\documentclass[aps,prd,nofootinbib]{revtex4-2}
\usepackage[colorlinks=true, pdfstartview=FitV, linkcolor=blue, citecolor=red, urlcolor=magenta]{hyperref}
\usepackage{graphicx}
\usepackage{latexsym}
\usepackage{amsmath}
\usepackage{amsfonts}
\usepackage{amssymb}
\usepackage{verbatim}
\usepackage{braket}

\def\p{\pi}

\newcommand{\be}{\begin{equation}}
\newcommand{\ee}{\end{equation}}
\newcommand{\bea}{\begin{eqnarray}}
\newcommand{\eea}{\end{eqnarray}}

\newcommand{\ben}{\begin{eqnarray}}
\newcommand{\een}{\end{eqnarray}}

\begin{document}

\title{Stefan–Boltzmann Law and Thermal Casimir Effect in Neutron Star Spacetime via Thermo Field Dynamics}

\author{$^{1}$K. E. L. de Farias}
\email{klecio.lima@uaf.ufcg.edu.br}

\author{$^{1,4}$M. A. Anacleto}
\email{anacleto@df.ufcg.edu.br}

\author{$^{1}$Rafael A. Batista}
\email{rafael.alves.batista@gmail.com}

\author{ $^{3}$Iver Brevik}
\email{iver.h.brevik@ntnu.no}

\author{$^{1,4}$F. A. Brito}
\email{fabrito@df.ufcg.edu.br}

\author{$^{1,4}$E. Passos}
\email{passos@df.ufcg.edu.br}

\author{$^{1}$Amilcar R. Queiroz}
\email{amilcarq@df.ufcg.edu.br}

\author{$^{1,2}$L\'azaro L. Sales}
\email{lazarolima@uern.br}


\affiliation{$^{1}$Departamento de F\'{\i}sica, Universidade Federal de Campina Grande,\\
Caixa Postal 10071, 58429-900, Campina Grande, Para\'{\i}ba, Brazil.}

\affiliation{$^{2}$Departamento de F\'{\i}sica, Universidade do Estado do Rio Grande do Norte,\\
59610-210, Mossor\'o-RN, Brazil}

\affiliation{$^{3}$Department of Energy and Process Engineering, Norwegian University of Science and Technology\\  N-7491 Trondheim, Norway.}

\affiliation{$^{4}$Unidade Acad\^emica de Matem\'atica, Universidade Federal de Campina Grande,\\ 58429-970,  Campina Grande, Para\'{\i}ba, Brazil.}



\begin{abstract}
We investigate the thermal Casimir effect for a massless scalar field in the curved spacetime of a neutron star within the Thermo Field Dynamics (TFD) formalism. Starting from the renormalized energy–momentum tensor, we generalize the Stefan–Boltzmann law to include gravitational redshift and curvature corrections governed by the Tolman–Oppenheimer–Volkoff (TOV) metric. Finite temperature and spatial compactification are introduced simultaneously, allowing a unified and consistent treatment of both vacuum and thermal contributions inside and outside the star. Analytical expressions are derived for the high- and low-temperature limits, showing explicitly how curvature and redshift modify the characteristic $T^4$ dependence of thermal radiation. The results reveal that strong gravity significantly alters the local energy density and pressure, demonstrating the nontrivial interplay between quantum vacuum fluctuations and compact astrophysical geometries. A polytropic model is considered to perform numerical analyses, highlighting the influence of the spacetime background on vacuum fluctuations.

\end{abstract}
 \maketitle


\section{Introduction}

The study of quantum vacuum fluctuations in curved spacetime provides a fundamental bridge between quantum field theory and gravity. One of the most striking manifestations of these fluctuations is the Casimir effect, first predicted by Casimir in 1948 \cite{Casimir:1948dh}, which arises from the modification of the zero-point energy of quantum fields due to boundary conditions and subsequently
modifications due to spacetime topology. Originally formulated in flat space and experimentally verified using distinct methods \cite{Lamoreaux:1996wh,Mohideen:1998iz,Bressi:2000iw}, the Casimir effect has since evolved into a powerful theoretical tool for investigating quantum phenomena.

In recent decades, the extension of the Casimir effect to curved spacetime has gained considerable attention \cite{birrell1984quantum,Cognola:1999zx,Lock:2016rmg,deFarias:2020xms}, as it reveals how curvature, topology, and boundary conditions can influence vacuum energy and tension \cite{Elizalde:2002dd,BezerradeMello:2011nv,Khanna:2011gf,Santos:2020taq,deFarias:2022rju}. In this context, quantum fields in the presence of strong gravitational fields, such as those near black holes \cite{Muniz:2015jba,Santos:2020taq} or within compact astrophysical objects \cite{Emig:2007cf,Milton:2014psa}, exhibit significant deviations from the behaviour of flat space. These deviations can affect both the local energy density and the vacuum pressure, and may play a role in the thermodynamic stability and radiation properties of compact stars.

Among these systems, neutron stars provide a particularly compelling environment in which quantum, thermal, and gravitational effects coexist under extreme and highly nonperturbative conditions. These compact remnants, formed during core-collapse supernovae, contain masses of order $\sim1-2\,M_\odot$ compressed into radii of $10-12\,\rm km$, leading to central densities that may exceed several times nuclear saturation density and curvature scales far beyond those accessible in any terrestrial experiment. The steep radial gradients of pressure, density, and gravitational redshift predicted by the Tolman–Oppenheimer–Volkoff (TOV) equations, \cite{Tolman:1939jz,Oppenheimer:1939ne}, create a natural arena for studying how strong gravity modifies the vacuum structure of quantum fields. As a result, neutron stars act as astrophysical laboratories where quantum vacuum fluctuations, finite-temperature effects, and boundary-induced stresses may exhibit behavior markedly distinct from that of flat spacetime as expected for a curved spacetime \cite{Harada:1998ge,Magierski:2001mg,Mota:2023ozt}. In particular, understanding how thermal radiation and Casimir-like vacuum stresses respond to the rapidly varying gravitational potential within neutron stars may offer insights into the deep interplay between quantum field theory in curved spacetime and relativistic astrophysics, with potential implications for the thermodynamics, stability, and radiative properties of compact stars.

To consistently introduce finite temperature effects into a curved spacetime background, we adopt the framework of Thermal Field Dynamics (TFD) \cite{Takahashi:1996zn,khanna2009thermal}. This real-time formalism allows the treatment of quantum fields at finite temperatures through the Bogoliubov transformation and Hilbert space doubling, thus enabling a natural description of thermal excitations in both flat and curved geometries. The TFD approach has been successfully applied to a variety of contexts \cite{Costa:2010fx,Leineker:2010uw,Cabral:2023xbf,Araujo:2024hnu,Irges:2025idm,Kalita:2025foa}, including the generalization of the Stefan-Boltzmann law and the evaluation of Casimir energies at finite temperatures \cite{Santos:2020bge,Santos:2020bge,Ulhoa:2023opw}.

Hence, the TFD formalism is used to investigate the thermal Casimir effect in the curved spacetime of a neutron star described by the Tolman–Oppenheimer–Volkoff (TOV) metric. The stellar interior is modeled through a relativistic polytropic equation of state \cite{chandrasekhar1957introduction,Nilsson:2000zg}, which provides a physically motivated and widely used description of neutron-star matter in terms of its pressure–density relation. This framework allows us to obtain self-consistent profiles for the gravitational potential, mass distribution, and redshift function throughout the star, all of which directly influence the structure of vacuum fluctuations. Starting from the action of a scalar field non-minimally coupled to gravity, we derive the renormalized energy–momentum tensor and analyze how the gravitational field, compactness, and the non-minimal coupling parameter that modifies the local vacuum energy density. By implementing temporal and spatial compactifications within the TFD approach, we obtain analytical expressions for the Casimir energy and pressure at finite temperature, unifying the flat, Schwarzschild, and neutron-star regimes under a single consistent formalism.

This work is organized as follows. In Sec. \ref{sec1}, a brief explanation is given about scalar field theory in curved spacetime. In the following Sec. \ref{sec2} it is presented the Neutron Star metric and we derive the Stefan-Boltzmann law, the renormalized Casimir energy and the radial Casimir pressure at null temperature modified by the NS spacetime and limit cases, i.e., Schwarzschild and Minkowski spacetimes. The finite temperature Casimir effect is studied in Sec. \ref{sec3}, where we find a closed expression for the Casimir energy and pressure and the low and high-temperature limits for each quantity. In Sec. \ref{sec4}, a detailed analysis is made by considering the polytropic model relative to the results obtained during this work. Finally, in Sec. \ref{sec5} we present the final comments. This work uses natural units for $\hslash=c=1$.

\section{Scalar Field Theory non-Minimally Coupled to Gravity}
\label{sec1}

The action describing a massless scalar field in a curved spacetime background is given by
\begin{equation}
S = \frac{1}{2} \int d^4x \, \sqrt{-g} \left( g^{\mu\nu} \nabla_{\mu} \phi(x) \nabla_{\nu} \phi(x) - \xi R \phi(x)^2 \right),
\label{eq1.1}
\end{equation}
where $g$ is the determinant of the metric tensor, $\xi$ is the coupling constant to gravity, $R$ is the Ricci scalar curvature, and $\nabla_{\mu}$ denotes the covariant derivative. For a massless field, choosing $\xi = \frac{1}{6}$ corresponds to the conformal coupling, which renders the theory invariant under conformal transformations.

The equation of motion for the scalar field is obtained by varying the action \eqref{eq1.1} with respect to $\phi(x)$, yielding
\begin{equation}
\left( \Box + \xi R \right) \phi(x) = 0,
\label{eq1.2}
\end{equation}
where $\Box = \nabla^{\mu} \nabla_{\mu}$ is the covariant d'Alembertian operator.

The energy-momentum tensor is defined as
\begin{equation}
T_{\gamma\rho} = -\frac{2}{\sqrt{-g}} \frac{\delta S}{\delta g^{\gamma\rho}}.
\label{eq1.3}
\end{equation}
Substituting the action \eqref{eq1.1} into the definition \eqref{eq1.3}, one obtains the explicit form of the energy–momentum tensor:
\begin{equation}
    T_{\gamma\rho} = \frac{1}{2} g_{\gamma\rho} g^{\mu\nu} \partial_{\mu} \phi(x) \partial_{\nu} \phi(x) -\partial_{\gamma} \phi(x) \partial_{\rho} \phi(x)  + \xi \left( R_{\gamma\rho} - \frac{1}{2} g_{\gamma\rho} R + g_{\gamma\rho} \Box - \nabla_{\gamma} \nabla_{\rho} \right) \phi(x)^2\,.
\label{eq1.4}
\end{equation}

The energy–momentum tensor plays a central role in the computation of thermodynamic and quantum effects, such as the Stefan–Boltzmann law and the Casimir effect. However, it gives a divergent result initially. To avoid this problem, we use a regularization  process, where one evaluates the tensor at separated spacetime points, leading to the point-split expression:
\begin{eqnarray}
T_{\gamma\rho}(x) =\lim_{x^{\prime} \to x} \tau \left[\frac{1}{2} g_{\gamma\rho} g^{\mu\nu} \partial_{\mu} \phi(x) \partial_{\nu} \phi(x^{\prime})  -\partial_{\gamma} \phi(x) \partial_{\rho} \phi(x^{\prime})
 + \xi \left( R_{\gamma\rho} - \frac{1}{2} g_{\gamma\rho} R + g_{\gamma\rho} \Box - \nabla_{\gamma} \nabla_{\rho} \right) \phi(x) \phi(x^{\prime}) \right],
\label{eq1.5}
\end{eqnarray}
where $\tau$ denotes the time-ordering operator.

Under canonical quantization, the equal-time commutation relation is given by
\begin{equation}
\left[ \phi(x), \partial^{\prime \mu} \phi(x^{\prime}) \right] = i n_0^{\mu} \delta^{(3)}(\vec{x} - \vec{x}^{\,\prime}),
\label{eq1.6}
\end{equation}
where $n_0^{\mu} = (1, 0, 0, 0)$ is a time-like unit vector. Additionally, the derivative of the Heaviside step function satisfies
\begin{equation}
\partial^{\rho} \theta(x^0 - x^{\prime 0}) = n_0^{\rho} \delta(x^0 - x^{\prime 0}).
\label{eq1.7}
\end{equation}

Rewriting equation \eqref{eq1.5} and using the commutation relation
\begin{equation}
\partial_{\gamma} \phi(x) \partial_{\rho} \phi(x^{\prime}) = \partial_{\gamma} \left[ \phi(x), \partial^{\prime \mu} \phi(x^{\prime}) \right] + \partial_{\gamma} \partial_{\rho} \left( \phi(x) \phi(x^{\prime}) \right),
\label{eq1.8}
\end{equation}
we obtain
\begin{align}
T_{\gamma\rho}(x)=& \lim_{x^{\prime} \rightarrow x} \left\{ \Gamma_{\gamma\rho} \, \tau[\phi(x)\phi(x^{\prime})] + \partial_{\rho} \left[ \theta(x^0 - x^{\prime 0}) \, i n_0^{\rho} \delta^{(3)}(\vec{x} - \vec{x}^{\,\prime}) \right]  + \frac{1}{2} g_{\gamma\rho} \partial^{\mu} \left( \theta(x^0 - x^{\prime 0}) \, i n_{0\mu} \delta^{(3)}(\vec{x} - \vec{x}^{\,\prime}) \right) \right\},
\label{eq1.9}
\end{align}
where $\theta(x^0 - x^{\prime 0})$ is the Heaviside step function and $\delta^{(3)}$ is the three-dimensional Dirac delta function.

Since $n_0^{\mu}$ is a purely time-like vector, only temporal derivatives contribute when acting on the step function, while the Dirac delta is spatial. Using the commutation relation \eqref{eq1.6}, the energy-momentum tensor simplifies to
\begin{equation}
T_{\gamma\rho}(x) = \lim_{x^{\prime} \rightarrow x} \left\{ \hat\Gamma_{\gamma\rho} \, \tau[\phi(x)\phi(x^{\prime})] -
I_{\gamma\rho} \delta^{(4)}(x - x^{\prime}) \right\},
\label{eq1.10}
\end{equation}
where we define the operator $\hat\Gamma_{\gamma\rho}$ as:
\begin{equation}
\hat\Gamma_{\gamma\rho}(x) = \frac{1}{2} g_{\gamma\rho} \partial^{\mu} \partial_{\mu}-\partial_{\gamma} \partial_{\rho}  + \xi \left( R_{\gamma\rho} - \frac{1}{2} g_{\gamma\rho} R + g_{\gamma\rho} \Box - \nabla_{\gamma} \nabla_{\rho} \right),
\label{eq1.11}
\end{equation}
and
\begin{equation}
I_{\gamma\rho} = -\frac{i}{2} g_{\gamma\rho} n_0^{\mu} n_{0\mu} + i n_{0\gamma} n_{0\rho}.
\label{eq1.12}
\end{equation}

The vacuum expectation value of the energy–momentum tensor, obtained from Eq. \eqref{eq1.10}, takes the form
\begin{equation}
\langle T_{\gamma\rho}(x) \rangle = \lim_{x^{\prime} \rightarrow x} \left\{ i \hat\Gamma_{\gamma\rho} G_0(x - x^{\prime}) - I_{\gamma\rho} \delta^{(4)}(x - x^{\prime}) \right\},
\label{eq1.13}
\end{equation}
where
\begin{equation}
i G_0(x - x^{\prime}) = \langle 0 | \tau[\phi(x)\phi(x^{\prime})] | 0 \rangle,
\label{eq1.14}
\end{equation}
is the Feynman propagator of the scalar field. This expression establishes a direct link between the geometry of spacetime and the quantum fluctuations of the field. The first term, involving $\Gamma_{\gamma\rho} G_0$, represents the contribution from vacuum correlations, while the second term subtracts local divergences that appear in the coincidence limit $x' \to x$.

By specifying the appropriate components of $\langle T_{\gamma\rho} \rangle$ and imposing suitable boundary or thermal conditions, one can derive a wide range of physical results. For instance, in flat spacetime at finite temperature, the $\langle T_{00} \rangle$ component reproduces the Stefan–Boltzmann law, whereas for confined geometries (e.g., between parallel plates or in curved backgrounds), it leads to the Casimir energy and pressure as we will see in the following section.

\section{The neutron star metric}
\label{sec2}

We consider the metric used in the Tolman–Oppenheimer–Volkoff (TOV) equations, which describes a static, spherically symmetric mass distribution in equilibrium, such as that found in neutron stars. The line element is given by
\begin{equation}
ds^2 = -e^{2\Phi(r)} dt^2 + \left(1 - \frac{2Gm(r)}{r}\right)^{-1} dr^2 + r^2 d\Omega^2,
\label{eq2.1}
\end{equation}
where $\Phi(r)$ is the gravitational potential, determined from the TOV equations, and $m(r)$ is the mass enclosed within a radius $r$, which satisfies
\begin{equation}
\frac{dm}{dr} = 4\pi r^2 \rho(r),
\label{eq2.2}
\end{equation}
with $\rho(r)$ being the energy density of the neutron star matter.

Based on the neutron star metric \eqref{eq2.1}, we can derive two differential operators from Eq.~\eqref{eq1.11}, which leads us to the time components $\gamma=\rho=00$ (see Apx. \ref{Apx1} for details.)
\begin{align}
\hat\Gamma_{00}= -\frac{1}{2} \partial_t\partial_{t^\prime} -e^{2\Phi}\left(\frac{1}{2}+\xi\right)\left[\left(1-\frac{2 G m}{r}\right) \partial_r\partial_{r^\prime}+\frac{1}{r^2}\left(\partial_{\theta}\partial_{\theta^\prime}+\frac{1}{\sin ^2 \theta} \partial_\phi\partial_{\phi^\prime}\right)\right] +\xi\frac{2 G e^{2 \Phi (r)} m'(r)}{r^2},
\label{eq2.10}
\end{align}
where it may be used to find the Stefan–Boltzmann law and the Casimir effect for a scalar field coupled to gravity by considering the appropriate boundary conditions, and taking the spatial components expressed by $\gamma=\rho=1$, we arrive at
\begin{align}
\hat\Gamma_{11}= & -\frac{1}{2} \partial_r\partial_{r^\prime} +\left(1-\frac{2 G m}{r}\right)^{-1}\left(\frac{1}{2}+\xi\right) \left[-e^{-2 \Phi} \partial_t\partial_{t^\prime}+ \frac{1}{r^2}\left(\partial_\theta\partial_{\theta^\prime}+\frac{1}{\sin ^2 \theta} \partial_\phi\partial_{\phi^\prime}\right)\right]\nonumber \\
& +2\xi\left\{\frac{r^2 \Phi '(r)-m(r)G \left[2 r \Phi '(r)+1\right]}{r^2 [r-2 G m(r)]}\right\}.
\label{eq2.11}
\end{align}

These operators, $\hat{\Gamma}_{00}$ and $\hat{\Gamma}_{11}$, constitute the basic ingredients for evaluating the vacuum expectation value $\langle T_{\mu\nu} \rangle$ at both zero and finite temperature, allowing the derivation of the Stefan-Boltzmann law and the Casimir energy and pressure in the neutron star background. Finite-temperature effects are consistently incorporated through the Thermo Field Dynamics (TFD) formalism, which provides
a real-time description of thermal fluctuations in curved spacetime.

It is worth emphasizing that the last term in each operator originates from the matter contribution to the curvature, where $R \neq 0$ inside the neutron star. These terms vanish in vacuum regions but remain finite in the stellar interior, thereby encoding the direct influence of the neutron-star matter distribution on both the generalized Stefan-Boltzmann law and the vacuum fluctuation structure of the scalar field.

\subsection{Stefan-Boltzmann law in Neutron Star Spacetime}

The introduction of temperature in Thermal Field Dynamics (TFD) is achieved by means of the Bogoliubov transformation together with the duplication of the Hilbert space, such that the thermal space is described by $S_T = S \otimes \tilde{S}$. Hence, the Bogoliubov transformation mixes the operators of $\tilde{S}$ and $S$, thereby introducing thermal effects through a thermal vacuum structure.

In order to calculate the Stefan-Boltzmann law, it is necessary to obtain the renormalized energy-momentum tensor, give by

\begin{equation}
\mathcal{T}_{\gamma\rho}(x;\alpha)=\braket{T^{AB}_{\gamma\rho}(x;\alpha)}-\braket{T^{AB}_{\gamma\rho}(x)}
\label{eq2.12}
\end{equation}
where the double-index notation $A,B$ arises from the TFD formalism, and $\alpha$ denotes a set of compactification parameters (such as the inverse temperature and compactification radii). Thus,
\begin{align}
\mathcal{T}_{\gamma\rho}(x;\alpha)=&\lim_{x^{\prime}\rightarrow x}\{i\Gamma_{\gamma\rho}[G^{AB}_0(x-x^{\prime};\alpha)-G^{AB}_0(x-x^{\prime})]\}\nonumber\\
=&\lim_{x^{\prime}\rightarrow x}\{i\Gamma_{\gamma\rho}\overline{G}^{AB}_0(x-x^{\prime};\alpha)\}.
\label{eq2.13}
\end{align}
where $\overline{G}^{AB}_0(x-x^{\prime};\alpha)$ is the renormalized propagator. The Fourier representation of the Green function is given by:
\begin{equation}
G^{AB}(x-x^{\prime};\alpha)=\frac{1}{(2\pi)^4}\int d^4ke^{ik(x-x^{\prime})}G^{AB}(k;\alpha),
\label{eq2.14}
\end{equation}
with
\begin{subequations}
    \begin{align}
G^{AB}(k;\alpha)=&B(\alpha)^{-1}G_0^{AB}(k)B(\alpha)\\
\overline{G}^{11}(k;\alpha)=&\overline{G}^{22}(k;\alpha)=v^2_k(\alpha)[G_0(k)-G_0^*(k)],
\label{eq2.15}
\end{align}
\end{subequations}
in which, $B(\alpha)$ is the Bogoliubov operator. Besides that
\begin{eqnarray}
G_0^{AB}(k)=\left(\begin{array}{cc}
G_0(k) & 0\\
0 & G_0^*(k)
\end{array}\right)=\left(\begin{array}{cc}
-\frac{1}{k^2+i\epsilon} & 0\\
0 & \frac{1}{k^2-i\epsilon}
\end{array}\right).
\label{eq2.16}
\end{eqnarray}
For the bosonic case, the Bogoliubov coefficient takes the general form:
\begin{equation}
v^2(k;\alpha)=\sum_{s=1}^d\sum_{\{\sigma_s\}}2^{s-1}exp\left[-\sum_{j=1}^{s}\alpha_{\sigma_j}l_{\sigma_j}k^{\sigma_j}\right].
\label{eq2.17}
\end{equation}
%
Then, from Eq.~\eqref{eq2.14}, we obtain
\begin{eqnarray}
G^{(11)}(x-x^{\prime};\alpha)&=&\frac{1}{(2\pi)^4}\int d^4k\left\{\sum_{s=1}^d\sum_{\{\sigma_s\}}2^{s-1}exp\left[-\sum_{j=1}^{s}\alpha_{\sigma_j}l_{\sigma_j}k^{\sigma_j}\right]\left[G_0(k)-G_0^*(k)\right]\right\}\nonumber\\
\ &=&2G_0(x-x^{\prime}-i\alpha l_{\sigma_0}n_0),
\label{eq2.18}
\end{eqnarray}
where $\alpha$ is a four-vector whose components are chosen according to the physical compactification.

To analyze the thermal case, we consider a massless scalar field coupled to a gravitational background. In general, the topology can be expressed as $\Gamma_D^d = \mathbb{S}^d \times \mathbb{R}^{D-d}$, where $D$ is the total number of dimensions and $d$ the number of compactified dimensions.

For the Stefan–Boltzmann case, we set $\alpha = (\beta, 0, 0, 0)$, with $\beta = 1/T$ representing the compactification of the time dimension. The resulting topology is $\Gamma_4^1 = \mathbb{S}^1 \times \mathbb{R}^3$. The physical components correspond to $A = B = 1$, and the generalized Bogoliubov transformation for a single compactified dimension is
\begin{equation}
v^2(\beta)=\sum_{\ell_0=1}^{\infty}exp\left[-\beta \ell_{0}k^{0}\right].
\label{eq2.19}
\end{equation}
Returning to renormalized energy-momentum tensor, Eq. \eqref{eq2.13}, we have
\begin{eqnarray}
\mathcal{T}_{\gamma\rho}(x;\beta)=\lim_{x^{\prime}\rightarrow x}\{i\hat\Gamma_{\gamma\rho}2\sum_{\ell_0=1}^{\infty}G_0(x-x^{\prime}-i\beta \ell_0n_0)\}\,.
\label{eq2.20}
\end{eqnarray}
The energy density associated with the system is obtained by choosing components $\alpha=\rho=0$, consequently
\begin{eqnarray}
\mathcal{T}_{00}(x;\beta)=2i\lim_{x^{\prime}\rightarrow x}\{\hat\Gamma_{00}\sum_{\ell_0=1}^{\infty}G_0(x-x^{\prime}-i\beta \ell_0n_0)\},
\label{eq2.21}
\end{eqnarray}
where the Green function is
\begin{eqnarray}
G_0(x-x^{\prime}-i\beta \ell_0n_0)=-\frac{i}{(2\pi)^2}\frac{1}{(x-x^{\prime}-i\beta \ell_0n_0)^2}\,,
\label{eq2.22}
\end{eqnarray}
and
\begin{equation}
\begin{aligned}
\left(x-x^{\prime}-i \beta \ell_0 n_0\right)^2 & =-e^{2\varphi(r)}\left(t-t^{\prime}-i \beta \ell_0\right)^2+\left(1-\frac{2 Gm(r)}{r}\right)^{-1}\left(r-r^{\prime}\right)^2 +r^2\left(\theta-\theta^{\prime}\right)^2+r^2 \sin ^2 \theta\left(\phi-\phi^{\prime}\right)^2
\end{aligned}
\label{eq2.23}
\end{equation}
then, the energy density at finite temperature $T$ reads
\begin{eqnarray}
\mathcal{T}_{00}(T)=-\frac{\pi ^2 e^{-2 \Phi (r)}
   \left(1+ \xi \right)}{30 \beta ^4}+\xi\frac{G m'(r)}{6 \beta ^2 r^2}\,.
\label{eq2.24}
\end{eqnarray}

The first term in the above equation reproduces the Stefan–Boltzmann law modified by the gravitational redshift factor $e^{-2\Phi(r)}$, while the second term represents the curvature correction associated with the interior mass distribution of the star where $m'(r)$ is the derivative of the mass function expressed by Eq. \eqref{eq2.2}, related to the energy density profile inside the neutron star. When the coupling constant $\xi$ vanishes, the curvature contribution disappears, recovering the minimal coupling case.

For an exterior Schwarzschild solution, describing a bounded sphere in vacuum, taking $e^{2\Phi(r)} = 1 - 2Gm/r$ in Eq.~\eqref{eq2.24} yields
\begin{eqnarray}
\mathcal{T}_{00}(T)=-\frac{\pi ^2
   \left(1+ \xi \right)}{30 \beta ^4g(r)}\,,
\label{eq2.25}
\end{eqnarray}
where $g(r)=1 - 2Gm/r$. This result, originally obtained in the context of static black hole thermodynamics (see Ref.~\cite{Santos:2020bge}), shows that the gravitational field enhances the local energy density of thermal radiation due to the redshift factor.

In the flat-space limit, taking $\Phi(r) \to 0$ and discarding the second terms of the r.h.s. of Eq.~\eqref{eq2.24} since $m'(r)=0$ out of the star, it is reduced to the well-known Stefan–Boltzmann law for a massless scalar field:
\begin{equation}
    \mathcal{T}_{00}(T) =-\frac{\pi^2(1+\xi)}{30\beta^4}\,,
    \label{eq2.26}
\end{equation}
which represents the standard blackbody energy density in the absence of curvature effects. The additional factor $(1+\xi)$ reflects the contribution of the non-minimal coupling between the scalar field and the curvature, which vanishes when $\xi=0$. The flat limit result \eqref{eq2.26} can also be obtained by taking $r \gg 2Gm$ from Eq. \eqref{eq2.25}.

Therefore, the result obtained in Eq.~\eqref{eq2.24} generalizes the Stefan–Boltzmann law to curved backgrounds, explicitly showing how the local gravitational potential $\Phi(r)$ and the internal mass distribution $m'(r)$ modify the thermal radiation inside compact astrophysical objects. These corrections become relevant in strong-gravity environments such as neutron stars, where the curvature can significantly alter the local energy density of quantum fields in thermal equilibrium. This establishes the foundation for the next sections, where spatial compactification and boundary effects will be incorporated to describe the Casimir contributions in curved spacetime.

Note that, although the exterior Schwarzschild limit still contains the term proportional to $\xi$, it no longer represents a coupling to the curvature, since $R=0$ in vacuum. In this regime, the term $\xi R\phi^2$ in Eq.~\eqref{eq1.1} vanishes, and $\xi$ simply acts as a parameter rescaling the normalization of the stress–energy tensor relative to the minimally coupled case. The same interpretation applies in the flat–spacetime limit.




\subsection{Casimir effect in the neutron star Spacetime}


Let us now consider the four-vector $\alpha = (0,i2a,0,0)$, which encodes the compactification along the radial spatial direction in the absence of temperature effects. In this configuration, the Casimir effect arises solely from the spatial compactification, with no contribution from thermal excitations. The constant \(a\) corresponds to the length of the compactified dimension, which determines the characteristic separation between the boundaries in the radial direction.

Within the thermo-field dynamics formalism, this choice of \(\alpha\) modifies the Bogoliubov transformation, which in this case takes the form
\begin{eqnarray}
v^2(\gamma) = \sum_{\ell_1=1}^{\infty} e^{-\alpha k^1 \ell_1},
\label{eq3.1}
\end{eqnarray}
where $\gamma$ is the parameter controlling the transformation, $k$ represents the momentum component along the compactified direction, and the summation index $\ell_1$ enumerates the discrete momentum modes arising from the spatial compactification.

The associated thermal doublet Green’s function is then written as
\begin{eqnarray}
G^{(11)}(x - x^{\prime}; \gamma) = G_0(x - x^{\prime} - i\gamma n_1 \ell_1)
\label{eq3.2}
\end{eqnarray}
where $G_0$ in the above equation is the zero-temperature, uncompactified Green’s function, and the argument shift by $-i\gamma n_1 \ell_1$ reflects the compactification in the second spatial coordinate. The unit vector $n_1$ indicates the direction of compactification, ensuring that only the radial coordinate is affected.

The Casimir energy density is obtained by evaluating the $00$-component of the energy--momentum tensor, $\mathcal{T}^{(11)}_{00}(x, a)$, for this configuration.  Following the standard procedure, one obtains the vacuum expectation value of the energy density associated with the spatial compactification, which yields:
\begin{align}
\mathcal{T}_{00}(x;a) = & \ 2i \lim_{x^{\prime} \rightarrow x} \left\{ \bar\Gamma_{00} \sum_{\ell_1=1}^{\infty} G_0\big(x - x^{\prime} - a n_1 \ell_1\big) \right\} \,,
\end{align}
consequently, the renormalized Casimir energy is given by
\begin{align}
\mathcal{E}_0^{\rm NS}(a)=& -\frac{e^{2 \Phi(r)} g_m(r)}{1440 \pi ^2 a ^4} \left\{\frac{180 a}{r}  (2 \xi +1) \zeta (3)M(r)+\pi^4(\xi+1)g_m(r)\right\}  - \xi \frac{G e^{2 \Phi (r)} g_m(r) m'(r)}{24 a ^2 r^2},
\label{eq3.3}
\end{align}
where $g_m(r) = g^{rr} = 1 - \frac{2Gm(r)}{r}$ is the radial component of the inverse metric and $M(r)=\frac{G m(r)}{r}-G m'(r)$. In the exterior region of the compact object, where $m'(r)=0$ (no additional matter distribution), the last term in Eq.~\eqref{eq3.3} vanishes, and the energy density simplifies to
\begin{align}
\mathcal{E}_0^{\rm BH}(a) = -\frac{g(r)^2}{1440 \pi ^2 a ^4} \left\{\frac{180 a  Gm}{r^2} (2 \xi +1) \zeta (3)  + \pi^4(\xi+1)g(r) \right\}
\label{eq3.4}
\end{align}
This expression describes the purely gravitationally modified Casimir effect outside the spherical body. The first term in the braces accounts for the combined effect of the compactification scale $a$ and the gravitational coupling $G$, while the second term corresponds to the purely kinematic contribution from the boundaries in curved spacetime.

Finally, taking the flat spacetime limit $g(r) \to 1$ $(r \gg 2Gm)$ removes all gravitational corrections, recovering the standard Casimir energy density for a massless scalar field:
\begin{equation}
    \mathcal{E}_0^{\rm Flat}(a) = -\frac{\pi^2(1+\xi)}{1440 a^4}\,,
    \label{eq3.5}
\end{equation}
which matches the well-known result in Minkowski space, following the additional factor $\xi$ accounting for the non-minimal coupling of the field.

It is worth highlighting that, as in the case of the modified Stefan–Boltzmann law, the factor $\xi$ remains present here, carrying the same physical meaning discussed earlier, that is, it parametrizes the deviation from minimal coupling and does not represent a direct interaction with the curvature in the Schwarzschild region.


When it is considered $\gamma=\rho=1$, the Casimir pressure at null temperature may be found, leading us to
\begin{align}
\mathcal{T}_{11}(x;a)= & \ 2i \lim_{x^{\prime} \rightarrow x} \left\{ \bar\Gamma_{11} \sum_{\ell_1=1}^{\infty} G_0\big(x - x^{\prime} - a n_1 \ell_1\big) \right\}\,,
\end{align}
then,
\begin{align}
\mathcal{P}_0^{\rm NS}(a)=&-\frac{\pi^2(1+\xi)g(r)}{480 a ^4 }-\frac{   M(r) \zeta (3)}{8 \pi ^2 a ^3 r}
+\frac{\xi  \left[ rg(r)\Phi'(r)-Gm(r)/r\right]}{24 a ^2 r^2}\,.
\label{eq3.6}
\end{align}

In the vacuum region outside the compact object, where $m'(r)=0$, the second term in the numerator of the first fraction becomes constant in $r $, and the derivative term involving $m'(r)$ in the $\xi$-correction vanishes. For the Schwarzschild metric,
\begin{equation}
    \Phi(r) = \frac{1}{2} \ln \left( 1 - \frac{2Gm}{r} \right),
    \label{eq3.7}
\end{equation}
so that
\begin{equation}
    \Phi'(r) = \frac{Gm}{r^2 g(r)}, \text{ with } g(r) = 1 - \frac{2Gm}{r}.
    \label{eq3.8}
\end{equation}
The radial pressure in this case simplifies to
\begin{equation}
    \mathcal{P}_0^{\rm BH}(a) = -\frac{\pi^2 (1 + \xi) g(r)}{480 a^4}-\frac{   \zeta (3) Gm}{8 \pi^2 a^3 r^2}.
    \label{eq3.9}
\end{equation}
This expression contains both the purely kinematic Casimir term (proportional to $\pi^2$) and gravitationally modified contributions proportional to $G$ and $\zeta (3)$.

Taking $G \to 0$ and $\Phi(r)\to 0$ in Eq. \eqref{eq3.6}, the gravitational terms disappear, and one recovers the standard Casimir radial pressure for a massless scalar field with curvature coupling:
\begin{equation}
    \mathcal{P}_0^{\rm Flat}(a) = -\frac{\pi^2(1 + \xi)}{480 \, a^4}.
    \label{eq3.10}
\end{equation}

The above expression gives us the Minkowski-space result for the pressure, modified only by the $(1 + \xi)$ factor due to the non-minimal coupling.


\section{Thermal Casimir effect in Neutron Star Spacetime}
\label{sec3}

To introduce the temperature in the Casimir effect, we choose the $\alpha=(\beta,i2a,0,0)$, where the presence of $\beta$ characterizes the introduction of the temperature, and $a$, as we have seen, is responsible for the Casimir effect. The Bogoliubov transformation is then

\begin{eqnarray}
v^2(\beta,a)=\sum_{\ell_0=1}^{\infty}e^{-\beta k^0\ell_0}+\sum_{\ell_1=1}^{\infty}e^{-a k^1\ell_1}+2\sum_{\ell_0,\ell_1=1}^{\infty}e^{-\beta k^0\ell_0-a k^1\ell_1}\,.
\label{eq4.1}
\end{eqnarray}
Then, the associated Green's functions will be
\begin{eqnarray}
G^{(11)}(\beta;\alpha)=G_0(x-x^{\prime}-i\beta n_0\ell_0)+G_0(x-x^{\prime}-a n_1\ell_1)+4G_0(x-x^{\prime}-i\beta n_0\ell_0-a n_1\ell_1)\,.
\label{eq4.2}
\end{eqnarray}

The first two terms have already been calculated, and their results are found in \eqref{eq2.24} and \eqref{eq3.3}. However, the last one is the term that provides the interaction between temperature and Casimir energy. Hence, the thermal Casimir energy is given by
\begin{align}
    \mathcal{E}^{\rm NS}(\beta;a)=&-\frac{2}{16\pi^2 a^4}\sum_{\ell_0,\ell_1=1}^{\infty}\left\{\frac{e^{2 \Phi (r)} g_m(r)}{  \left[\gamma_r ^2 \ell_0^2 + \ell_1^2 \right]^3}   \left\{\ell_1^2  \left[(\xi +1)  g_m(r)+(2 \xi +1) \frac{2 a \ell_1}{r}M(r) \right]\right.\right.\nonumber\\
   &\left.\left.-\gamma_r ^2 \ell_0^2 \left[(2 \xi +1)\frac{2 a  \ell_1}{r}  \left[2 r g_m(r) \Phi '(r)+M(r)\right]+3 (\xi +1)  g_m(r)\right] \right\} -\frac{8a^2G \xi  e^{2 \Phi (r)}g_m(r) m'(r)}{ r^2 \left[\gamma_r ^2 \ell_0^2 +\ell_1^2\right]}\right\}\,,
   \label{1.38}
\end{align}
where we define the dimensionless parameters $\gamma_r^2=\beta^2e^{2 \Phi (r)}g_m(r)/(2a)^2$. The expression in Eq.~\eqref{1.38} represents the most general form of the thermal Casimir energy density in the curved spacetime of a neutron star, including both the vacuum curvature effects and the finite-temperature corrections. The double summation over $\ell_0$ and $\ell_1$ accounts for the combined effects of temporal periodicity (temperature) and spatial compactification (Casimir geometry).

In the Schwarzschild limit, the geometry simplifies, and the metric functions become explicit functions of $r$ through $g(r) = 1 - 2Gm/r$. This leads to:
\begin{align}
    \mathcal{E}^{\rm BH}(\beta;a)=&\frac{1}{8\pi^2a^4}\sum_{\ell_0,\ell_1=1}^{\infty}\left\{\frac{ g(r)^2 [ (\xi +1) g(r) + 2(2 \xi +1) a  G \ell_1 m/r^2 ] \left[3 (\gamma_r\ell_0)^2- \ell_1^2 \right]}{\left[(\gamma_b \ell_0 )^2+\ell_1 ^2\right]^3}\right\}\,,
   \label{1.39}
\end{align}
and $\gamma_b=\beta g(r)/(2a)$. The expression still retains curvature-dependent modifications to the thermal Casimir energy, but with a simpler dependence on $m$ and $r$. The coupling parameter $\xi$ continues to play a role in modifying both the magnitude and sign of the effect, potentially leading to suppression or enhancement depending on the geometry and temperature.

In the flat-spacetime limit, the expression reduces to:

\begin{align}
    \mathcal{E}^{\rm Flat}(\beta;a)=&\frac{1}{8\pi^2a^4}\sum_{\ell_0,\ell_1=1}^{\infty}\left\{\frac{ (\xi +1) \left[3( \gamma_r \ell_0)^2-\ell_1^2\right]}{ \left[(\gamma_p \ell_0)^2 +\ell_1^2\right]^3} \right\}\,,
   \label{1.40}
\end{align}
where $\gamma_p=\beta/(2a)$. Eq.~\eqref{1.40} recovers the standard flat-spacetime result for the thermal Casimir effect of a scalar field. In this case, the dependence on $\beta$ and $a$ becomes symmetric under interchange of temporal and spatial compactification lengths, consistent with the well-known duality in finite-temperature field theory. This limit provides a natural check of the curved-space formalism, ensuring consistency with the well-established flat-space results.



The pressure expressed in terms of the temperature can be written as
\begin{align}
    \mathcal{P}^{\rm NS}(\beta;a)=&\frac{1}{8\p^2 a^4}\sum_{\ell_0,\ell_1=1}^{\infty} \left\{\frac{4a^2\xi[G m(r)- r^2g_m(r) \Phi '(r)]}{ r^3\left[(\gamma_r \ell_0)^2 + \ell_1^2 \right]}-\frac{1}{ \left[(\gamma_r \ell_0)^2
    +\ell_1^2\right]^3} \left[\frac{2a \ell_1^3M(r)}{ r} \right.\right.\nonumber\\
   &\left. \left.
    -  \frac{  (2\gamma_r \ell_0)^2 \ell_1a}{r} \left[ r g_m(r) \Phi '(r)+\frac{M(r)}{2}
   \right]- (\xi +1)g_m(r) \left[(\gamma_r \ell_0)^2-3 \ell_1^2\right]\right]\right\}.
   \label{1.41}
\end{align}
This expression encodes the interplay between thermal contributions (via the Matsubara index $\ell_0$), the compactification scale $\alpha$ (through $\ell_1$), and the background geometry, represented by the mass function $m(r)$, the metric factor $g(r)$, and the redshift potential $\Phi(r)$. The presence of the coupling $\xi$ also allows one to interpolate between minimal and conformal coupling regimes.

In order to check the consistency of this general result, it is instructive to analyze some particular limits of physical interest.

For a Schwarzschild limit, the pressure obtained for a neutron star background, Eq. \eqref{1.41}, is reduced to
\begin{align}
    \mathcal{P}^{\rm BH}(\beta;a)=\frac{1}{8\pi^2a^4}&\sum_{\ell_0,\ell_1=1}^{\infty}\frac{1}{\left[(\gamma_b \ell_0 )^2+ \ell_1^2
   \right]^3}\left\{(\xi +1) g(r) \left[(\gamma_b \ell_0 )^2-3\ell_1 ^2\right]+ \frac{2 a  G \ell_1 m \left[3 (\gamma_b \ell_0 )^2 - \ell_1^2\right]}{r^2}  \right\}\,.
   \label{1.42}
\end{align}
This form highlights the explicit dependence on the black hole mass $m$ and the Schwarzschild factor $g(r)$, confirming that the pressure is strongly suppressed near the event horizon due to gravitational redshift effects.

In the limit of flat spacetime, the expression is written as
\begin{equation}
    \mathcal{P}^{\rm Flat}(\beta;a)=\frac{1}{8\p^2a^4}\sum_{\ell_0,\ell_1=1}^{\infty}\frac{ (\xi +1) \left[(\gamma_p \ell_0)^2-3 \ell_1^2\right]}{\left[(\gamma_p \ell_0)^2+
   \ell_1^2\right]^3}\,.
   \label{1.43}
\end{equation}
This result reproduces the well-known flat-space expression for the thermal Casimir pressure. In this limit, the purely geometric contributions vanish, and the dependence on $\xi$ becomes transparent: the conformally coupled case $\xi = 1/6$ exhibits milder divergences compared to the minimally coupled case $\xi=0$.


\subsection{High temperature limit}

From the previously obtained expressions for the thermal Casimir energy and pressure, it is important to emphasize that the double summation over the indices $\ell_0$ and $\ell_1$ cannot be performed simultaneously, as it leads to divergent results. However, evaluating the sums in different orders provides useful approximations that reveal the asymptotic structure of the system and help identify the regime in which the physical quantities exhibit well-behaved temperature dependence.

In the high-temperature regime ($\gamma_r \to 0$, equivalently $k_{\rm B}T(2a)^2/e^{2 \Phi (r)}g_m(r) \to \infty$), the Casimir energy density presents a markedly distinct behaviour from that in the low-temperature limit. To analyze this case, we first perform the summation over the Matsubara frequency index $\ell_0$. After this procedure, the Casimir energy density in the high-temperature limit takes the following form:

\begin{align}
   \mathcal{E}_{\rm high}^{\rm NS}(\beta;a)= &g_m(r) e^{2 \Phi (r)}\sum_{\ell_1=1}^{\infty}\left\{\frac{ \left[ 2 a \ell_1 (2 \xi +1) M(r)+(\xi +1) r g_m(r)\right]}{16 \pi
   ^2 a^4  \ell_1^4 r} -\frac{G \xi  m'(r)}{2 \pi ^2 a^2 \ell_1^2 r^2}\right.\nonumber\\
   &-\frac{ \pi ^3 \ell_1^3 \coth \!\left(\tfrac{\pi  l_1}{\gamma_r}\right)
   \text{csch}^2\!\left(\tfrac{\pi  \ell_1}{\gamma_r}\right) \left[a \ell_1 (2 \xi +1)\Psi_{+}(r) +r g_m(r)(\xi +1)\right]}{16 \pi
   ^2 a^4 \gamma_r^3 \ell_1^4 r}\nonumber\\
   &\left.+\frac{(2 \xi +1)  \left[\gamma_r \coth \!\left(\tfrac{\pi  \ell_1}{\gamma_r}\right)+\pi  \ell_1 \text{csch}^2\!\left(\tfrac{\pi  \ell_1}{\gamma_r}\right)\right] \Psi_{-}(r)}{32 \pi  a^3 \gamma_r^2 \ell_1^2 r} +\frac{G \xi \coth\left(\tfrac{\pi  \ell_1}{\gamma_r}\right) m'(r)}{\pi a^2   \gamma_r \ell_1 r^2}\right\},
   \label{h1}
\end{align}
where,
\begin{equation}
    \Psi_{\pm}(r) =r g_m(r) \Phi '(r)\pm M(r)
    \label{h2}
\end{equation}

Considering asymptotic behaviour of the hyperbolic functions in eq,\eqref{h1}, the series can be evaluated for $l_1$, and the Casimir energy density can be expressed in terms of special functions and logarithmic contributions as
\begin{align}
   \mathcal{E}_{\rm high}^{\rm NS}(\beta;a)\approx &\mathcal{E}_{0}^{\rm NS}(a)+ g_m(r) e^{2 \Phi (r)}\left\{\frac{3G \xi   m'(r)}{24  a^2  r^2}+\frac{(2 \xi +1)}{2a^3r} \left[\frac{\zeta(3)}{\pi^2}-\frac{L(\gamma_r)}{4
   \gamma_r^2 } \right] \Psi_{-}(r)\right.\nonumber\\
   &\left.+\frac{\pi }{8 a^4 \gamma_r^3}  \left[2 (\xi +1) g_m(r)L(\gamma_r)-\frac{a (2 \xi +1) \left[\coth \!\left(\tfrac{\pi }{\gamma_r}\right)-1\right] \Psi_{+}(r)}{r}\right]
   \right\}\,,
   \label{h3}
\end{align}
where, $L(\gamma_r)=\ln \!\left(1-e^{-\tfrac{2 \pi }{\gamma_r}}\right)$.

$ \frac{2a T}{e^{\Phi (r)}g_m(r)^{1/2}}$
Note that there is also one part that is temperature-independent, which is the same as that obtained in the Casimir energy given by Eq. \eqref{eq3.3}. As it was already analysed earlier, and our interest is in the thermal part, we can ignore it. Considering only the thermal part, it diverges at low-temperature as $\gamma_r\to\infty\,(k_{\rm B}T(2a)^2/e^{2 \Phi (r)}g_m(r)\to0)$, which shows the impossibility to obtain both results

In the case of a black hole geometry, the Casimir energy density is modified by curvature effects. It takes the form
\begin{align}
    \mathcal{E}_{\rm high}^{\rm BH}(\beta;a)=&\mathcal{E}_{0}^{\rm BH}(a)+\frac{g(r)^2}{ \pi ^2 a^4 \gamma_b^3 } \left[ {\pi ^3 g(r)(\xi +1)}
    L(\gamma_b)-\frac{ a G m (2 \xi +1)  \pi ^3 \left[\coth \!\left(\tfrac{\pi }{\gamma_b}\right)-1\right]}{r^2} \right].
    \label{h4}
\end{align}
Here, the dependence on $\gamma_r^{-3}$ reflects the strong enhancement of thermal fluctuations in the high-temperature regime. The logarithmic corrections are also more pronounced, and the curvature-dependent terms involving $M(r)$ and $m(r)$ introduce characteristic deviations from the flat case.

For the Minkowski background, the high-temperature Casimir energy simplifies to
\begin{equation}
   \mathcal{E}_{\rm high}^{\rm Flat}(\beta;a)= \mathcal{E}_{0}^{\rm Flat}(a) +\frac{\pi  (\xi +1)}{4 a^4 \gamma_p^3} L(\gamma_p),
   \label{h5}
\end{equation}
where the first term corresponds to the dominant Stefan–Boltzmann-like contribution, while the second term encodes thermal corrections due to the compactification scale $a$ and the parameter $\gamma_r$.

In summary, at high temperature, the Casimir energy is dominated by thermal excitations, with a leading $\gamma_r^{-3}$ dependence that signals the breakdown of exponential suppression. Curvature introduces additional modifications in the black hole case, while the flat-space result recovers the expected thermodynamic behavior of quantum fields at high temperature.

For the neutron star geometry, the high-temperature limit of the Casimir pressure reads

\begin{align}
\mathcal{P}_{\rm high}^{\rm NS}(\beta;a)=&-\mathcal{P}_{0}^{\rm NS}(a) +\frac{\xi  r g_m(r) \Phi '(r)-G \xi  m(r)/r}{12 a^2 r^2}
-\frac{ K(\gamma_r)}{64 a^3r \gamma_r^3}\left[M(r)-\frac{rg(r)\Phi'(r)}{3}\right]\nonumber\\
   &+\frac{(\xi +1) g(r)}{16 \pi  a^4 \gamma_r^3} \left[-4 \pi  \gamma_r
   \text{Li}_2\left(e^{-\frac{2 \pi }{\gamma_r}}\right)+4 \pi ^2 L(\gamma_r)-
   \zeta (3) \gamma_r^2\right] \nonumber\\
   &+\frac{M(r) }{128 a^3 r\gamma_r^3}\left\{\gamma_r
   \left[\pi  \gamma_r-24  L(\gamma_r)\right]+8 \pi \left[\coth
   \left(\frac{\pi }{\gamma_r}\right)-1\right]\right\}
   \label{h6}
\end{align}
where,
\begin{equation}
    K(\gamma_r)=-\pi  \gamma_r^2+ 24 \gamma_r L(\gamma_r)+24 \pi  \left[\coth \left(\frac{\pi
   }{\gamma_r}\right)-1\right].\nonumber
   \label{h7}
\end{equation}

For the Schwarzschild-like black hole geometry, the pressure takes the form
\begin{align}
    \mathcal{P}_{\rm high}^{\rm BH}(\beta;a)=&-\mathcal{P}_{0}^{\rm BH}(a)-\frac{ K(\gamma_b)}{64 a^3 r \gamma_b^3}\frac{Gm}{r} +\frac{(\xi +1) g(r)}{16 \pi  a^4 \gamma_b^3} \left[-4 \pi  \gamma_b
   \text{Li}_2\left(e^{-\frac{2 \pi }{\gamma_b}}\right)+4 \pi ^2 L(\gamma_r)-
   \zeta (3) \gamma_b^2\right]\nonumber\\
   &+\frac{  G m}{128 a^3 r^2 \gamma_b^3}\left\{\gamma_b[\pi\gamma_b-24 L(\gamma_b)]+8\pi  \left[\coth \left(\frac{\pi }{\gamma_b}\right)-1\right]\right\}
   \label{h8}
\end{align}

In the flat-space limit, all curvature-dependent terms vanish, and the high-temperature Casimir pressure reduces to the well-known thermodynamic form
\begin{equation}
    \mathcal{P}_{\rm high}^{\rm Flat}(\beta;a)=-\mathcal{P}_{0}^{\rm Flat}(a) +\frac{(\xi +1)}{16 \pi  a^4 \gamma_p^3} \left[-4 \pi  \gamma_p \text{Li}_2\left(e^{-\frac{2 \pi }{\gamma_p}}\right) +4 \pi ^2 L(\gamma_p)-3 \zeta (3) \gamma_p^2\right]\,.
    \label{h9}
\end{equation}
This result recovers the expected high-temperature behaviour of quantum fields between parallel plates, with the pressure dominated by the thermal part of the spectrum.


\subsection{Low temperature limit}

In the regime of low temperature ($\gamma_r \to \infty$, equivalently $k_{\rm B}T(2a)^2/e^{2 \Phi (r)}g_m(r) \to 0$), the thermal corrections to the Casimir energy become exponentially suppressed. This behaviour arises when the summation over $l_1$ in Eq.~\eqref{1.38} is performed first, followed by an expansion for large $\gamma_r$. The resulting expression for the energy density reads
\begin{align}
     \mathcal{E}_{\rm low}^{\rm NS}(\beta;a)\approx&g(r) e^{2 \Phi (r)} \left\{\frac{  G \xi  m'(r)}{6 a^2 r^2\gamma_r^2}- \frac{\pi  (\xi +1) g(r) }{480 \pi ^2 a^4\gamma_r^4} \left[\pi ^2 \left(120 \gamma_r^3 L(\gamma_r^{-1})+\pi \right)-30 \gamma_r \left(4 \pi  \gamma_r \text{Li}_2\!\left(e^{-2 \pi
    \gamma_r}\right)+\zeta (3)\right)\right]\right\},
    \label{l1}
\end{align}
which clearly shows that the dominant contributions at low temperature are controlled by inverse powers of $\gamma_r$, reflecting the suppression of thermal fluctuations. It is straightforward that in the limit where $a\to0$, the energy density diverges, while it vanishes as $a\to\infty$. Moreover, Eq. \eqref{1.38} also vanishes as $\gamma_r\to0$ as expected. In particular, the first term proportional to $m'(r)$ encodes the effect of spacetime curvature, while the second captures the interplay between vacuum and thermal corrections.

For the Schwarzschild case, the low-temperature Casimir energy density takes the simplified form
\begin{equation}
  \mathcal{E}_{\rm low}^{\rm BH}(\beta;a)= -\frac{g(r)^3  (\xi +1)  }{480 \pi a^4\gamma_b^4} \left\{\pi ^2 \left[120 \gamma_b^3
   L(\gamma_b^{-1})+\pi \right]-30 \gamma_b \left[4 \pi \gamma_b \text{Li}_2\!\left(e^{-2 \pi  \gamma_b}\right)+\zeta
   (3)\right]\right\},
   \label{l2}
\end{equation}
where the factor $g(r)^3$ strongly suppresses the energy near the event horizon. The curvature of spacetime amplifies the decay of thermal contributions, effectively enhancing the dominance of the vacuum part at sufficiently low temperatures.

In Minkowski spacetime, the expression reduces to
\begin{equation}
    \mathcal{E}_{\rm low}^{\rm Flat}(\beta;a)=-\frac{(\xi +1)}{480 \pi  a^4 \gamma_p ^4} \left\{\pi ^2 \left[120 \gamma_p ^3 L(\gamma_p^{-1})+\pi \right]-30 \gamma_p  \left[4 \pi  \gamma_p  \text{Li}_2\!\left(e^{-2 \pi  \gamma_p }\right)+\zeta
   (3)\right]\right\},
   \label{l3}
\end{equation}
which recovers the expected flat-space Casimir behaviour in the low-temperature limit. Here the $\gamma_r^{-4}$ dependence emphasizes the rapid suppression of the thermal part, leaving the leading term as the zero-temperature Casimir energy density.

The pressure in the neutron star at low temperature is given by
\begin{align}
    \mathcal{P}_{\rm low}^{\rm NS}(\beta;a)= &-\frac{\pi(\xi +1) g_m(r) \left[ 360 \gamma_r ^3 L(\gamma_b^{-1})+\pi \right]}{1440
   a^4 \gamma_r ^4} -\frac{ \xi  \left[G m(r)-r^2 g(r) \Phi '(r)\right]}{24 \pi  a^2 \gamma_r ^2 r^3}\nonumber\\
   &-\frac{M(r) \left[\gamma_r^3+3 \pi ^2 \gamma_r^2 \text{Li}_3\left(e^{-2 \pi  \gamma_r}\right)+3 \pi  \text{Li}_5\left(e^{-2 \pi  \gamma_r}\right)\right]}{96 a^3 \gamma_r^5 r}\nonumber\\
   &+\frac{e^{2 \Phi(r)}g_m(r)\beta^2}{ar}\left[-\frac{3 \zeta (3) }{4 \pi ^2 \gamma_r^3}+ \pi   \coth (\pi \gamma_r)-\pi  \right]\frac{\left[2 r g_m(r) \Phi'(r)+M(r)\right]}{64a^4\gamma_r^2}\,,
   \label{l4}
\end{align}
for a neutron-star spacetime, where both the local curvature and the gravitational potential $\Phi(r)$ contribute to the Casimir pressure. The first term represents the standard vacuum contribution, while the remaining terms encode curvature corrections arising from the radial dependence
of the metric functions $g(r)$, $\Phi'(r)$, and the mass profile $M(r)$. These geometric contributions can either enhance or suppress the pressure depending on the interplay between gravitational redshift and local curvature. Consequently, the Casimir pressure inside the star reflects not only boundary effects but also the strong-gravity environment, which modifies the thermal and vacuum fluctuations in a nontrivial way.

For the black-hole case, the pressure becomes

\begin{align}
    \mathcal{P}_{\rm low}^{\rm BH}(\beta;a)= &\frac{-\pi (\xi +1) g(r) \left[360 \gamma_b^3 \log \left(1-e^{-2 \pi  \gamma_b}\right)+\pi \right]}{1440  a^4
   \gamma_b^4} +\frac{\left[18 \pi  \gamma_b^2 \coth (\pi  \gamma_b)-18 \pi  \gamma_b^2-1\right]}{96 a^3 \gamma_b^2 r}\frac{Gm}{r}\nonumber\\
   &-\frac{\pi \left[\pi  \gamma_b^2 \text{Li}_3\left(e^{-2 \pi  \gamma_b}\right)+\text{Li}_5\left(e^{-2 \pi  \gamma_b}\right)\right]}{32 a^3
   \gamma_b^5 r}\frac{Gm}{r}-\frac{9 G m \zeta (3)}{64 \pi ^2 a^3 \gamma_b^3 r^2}
   \label{l5}
\end{align}
indicating that near the event horizon, the pressure is further suppressed by the factor $g(r)$, consistent with the redshifted thermal spectrum in strong gravity.

In flat spacetime, the pressure takes the simpler form
\begin{equation}
    \mathcal{P}_{\rm low}^{\rm Flat}(\beta;a)= -\frac{\pi  (\xi +1) \left(360 \gamma_p^3 \log \left(1-e^{-2 \pi  \gamma_p}\right)+\pi \right)}{1440 a^4 \gamma_p^4}
    \label{l6}
\end{equation}
which reproduces the expected negative Casimir pressure, driving an attractive force between the plates.

Overall, these results confirm that at low temperature the Casimir energy is dominated by its vacuum component, with thermal corrections exponentially small in $\gamma_r$. The presence of curvature-dependent terms introduces further modifications in the neutron star and black hole cases, but the qualitative feature of strong suppression of thermal effects remains universal.


\section{TOV equation and the Polytropic Equation of State}
\label{sec4}

The results obtained in the previous sections are expressed in terms of the TOV equations. We now analyze how these equations modify the Stefan–Boltzmann law and the Casimir effect at both zero and finite temperatures. As a first step, we must specify the equation of state (EoS) to be used. Since we are considering the simplest model of a neutron star, we adopt a polytropic EoS, which provides a straightforward yet effective description of stellar matter. A polytropic equation of state is defined by
\begin{equation}
P = K\,\rho^{\Gamma}\,,
\label{eqB2.13}
\end{equation}
with constants $K>0$ and adiabatic index $\Gamma>1$. Considering a barotropic fluid, the first law leads us to
\begin{equation}
d\!\left(\frac{\epsilon}{\rho}\right) = -\,P\,d\!\left(\frac{1}{\rho}\right),
\label{eqB2.14}
\end{equation}
which integrates (taking $\epsilon\to \rho$ as $P\to 0$) to
\begin{equation}
\epsilon(\rho) = \rho + \frac{P}{\Gamma-1}
= \rho + \frac{K}{\Gamma-1}\,\rho^{\Gamma}.
\label{eqB2.15}
\end{equation}

The relations above, together with the TOV equations presented in Appendix~\ref{Apx2}, provide the mathematical framework for the numerical implementation\footnote{https://github.com/niksterg/pyTOV/blob/master/TOV-polytropes.ipynb} required to investigate the behaviour of the modified Stefan–Boltzmann law in the curved spacetime of a neutron star.
\begin{figure}[h!]
    \centering
    \includegraphics[width=0.55\linewidth]{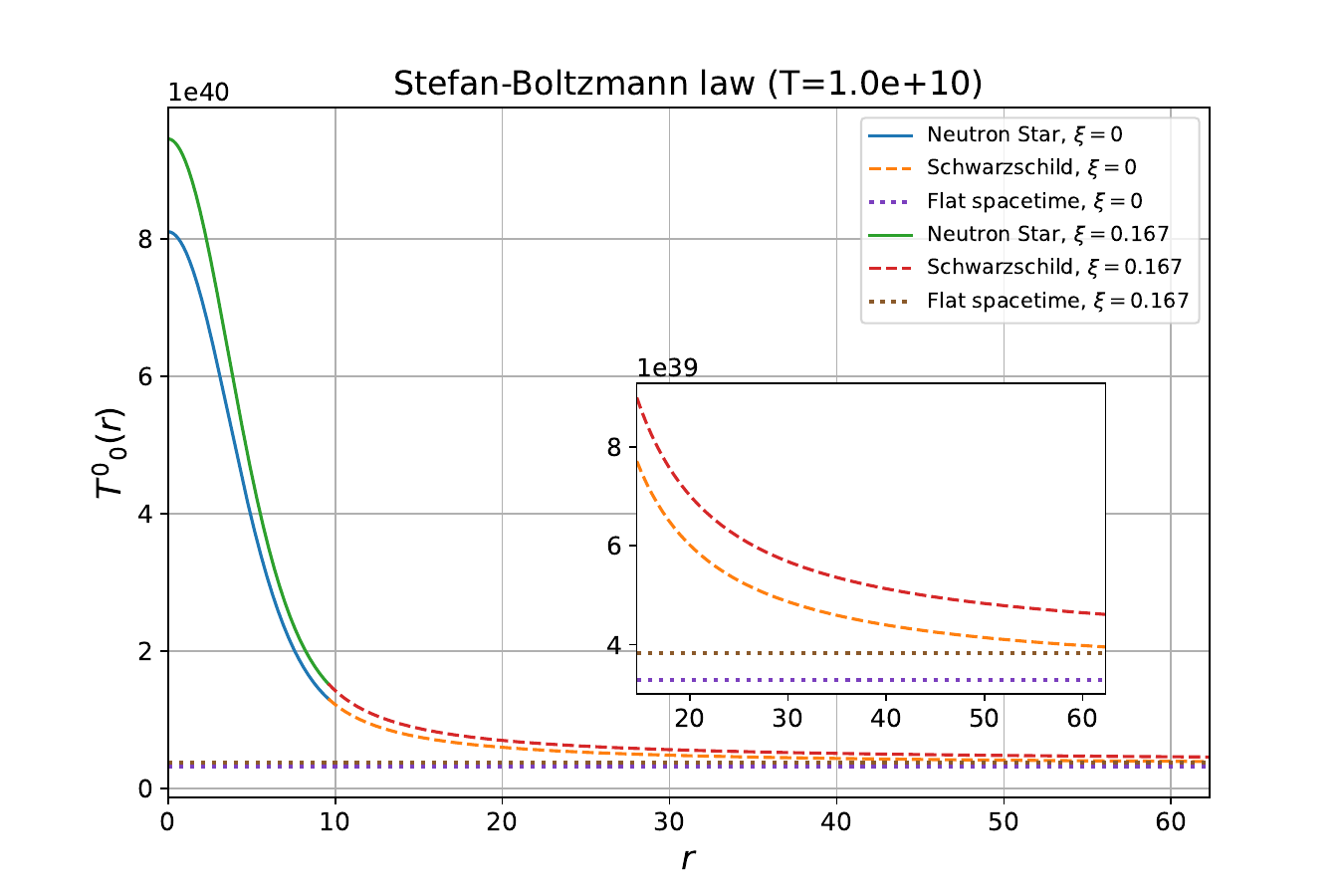}
    \caption{Radial profile of the energy-momentum tensor component $T^0_0(r)$ for a massless scalar field in thermal equilibrium, plotted for both the neutron star interior (solid lines) and the exterior Schwarzschild region (dashed lines) by a fixed temperature $T$, considering the minimal coupling $(\xi=0)$ and conformal coupling $(\xi=1 / 6)$. The dotted lines correspond to the flat spacetime Stefan-Boltzmann limit.}
    \label{nstar}
\end{figure}

Fig. \ref{nstar} shows the radial dependence of the covariant energy-momentum tensor component $T^0_{\ \ 0}(r)$, which represents the local radiative energy flux of a massless scalar field at finite temperature in several conditions of spacetime from Eqs. \eqref{eq2.24}--\eqref{eq2.26}. Inside the neutron star (solid lines), the Stefan--Boltzmann law increases toward the center due to the gravitational redshift encoded in $e^{\nu(r)}$, while in the exterior Schwarzschild region (dashed lines) the same redshift factor governs the thermal distribution, leading the curves to asymptotically approach the flat-space Stefan-Boltzmann value (dotted lines).
The inclusion of the coupling parameter $\xi$ modifies the overall normalization by a factor of $(1+\xi)$, with $\xi=1 / 6$ corresponding to the conformally coupled case and allows the influence of a correction term obtained in Eq. \eqref{eq2.24} as already mentioned earlier.
As expected, the neutron-star profile continuously matches the Schwarzschild solution at the surface, and converges to the flat spacetime when $r\to\infty$.

\begin{figure}[h!]
    \centering
    \includegraphics[width=0.45\linewidth]{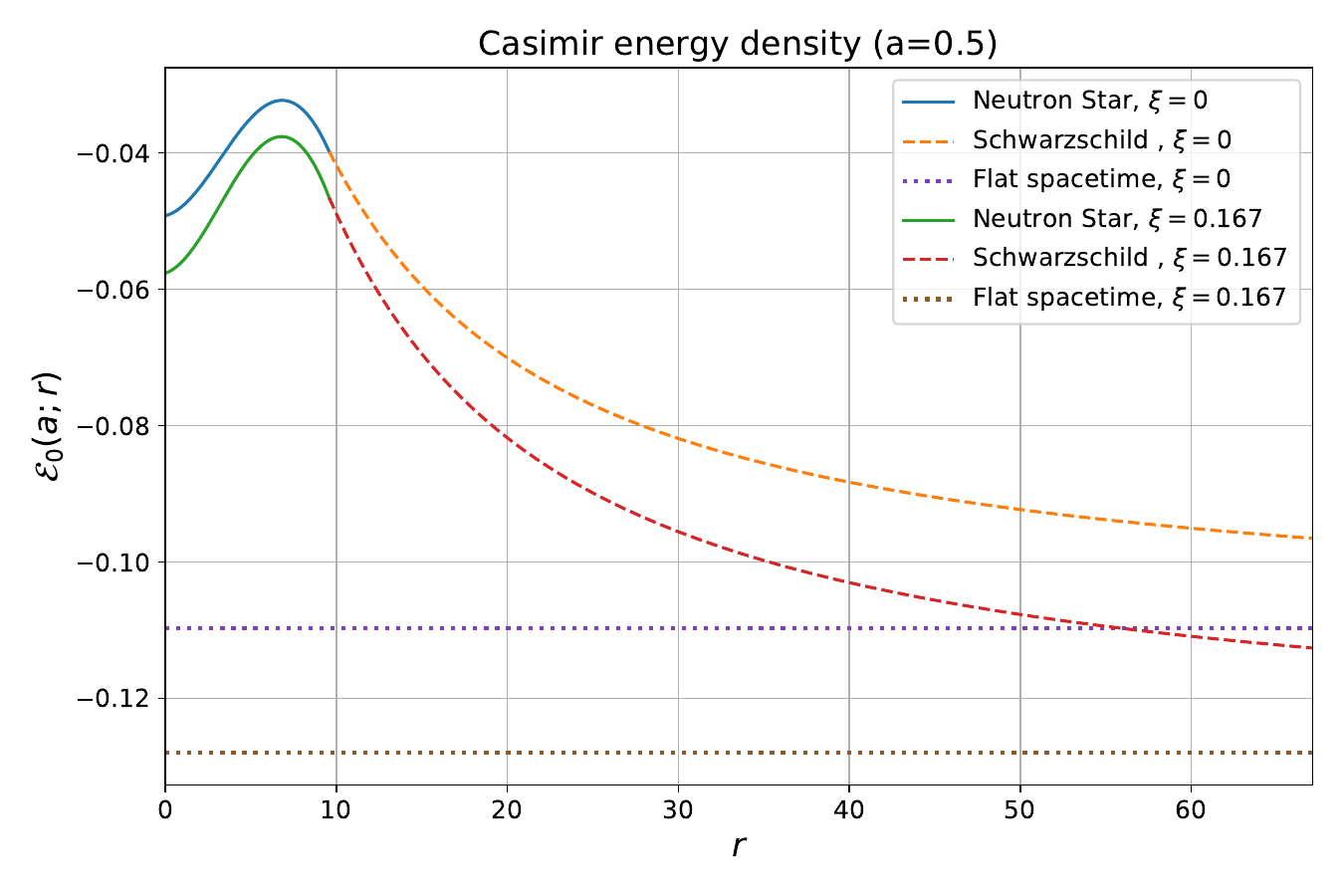}
    \includegraphics[width=0.45\linewidth]{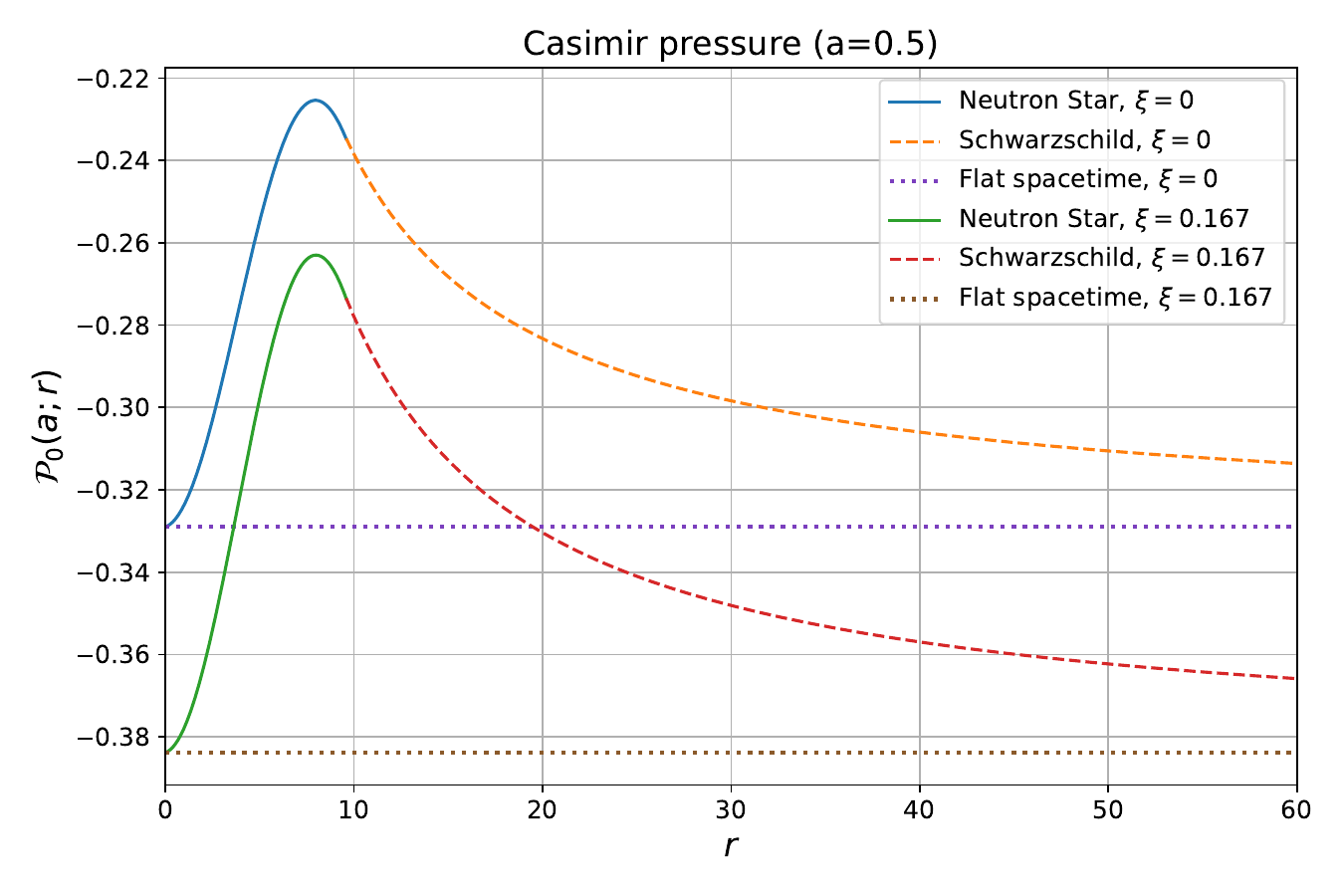}
    \caption{The left panel shows the renormalized Casimir energy density $\mathcal{E}_0(a;r)$ for a massless scalar field inside and outside a neutron star, with plate separation $a = 0.5$. The right panel shows the behaviour of the Casimir pressure with $a=0.5$ by considering the minimal coupling case ($\xi = 0$), the conformal coupling ($\xi = 1/6$) and the flat spacetime result for each case.}
    \label{caszero}
\end{figure}

In the case of the zero temperature renormalized Casimir energy expressed by Eqs. \eqref{eq3.3}--\eqref{eq3.5} and the Casimir pressure, Eqs. \eqref{eq3.6}--\eqref{eq3.8}, the behaviours of each are demonstrated by Figs. \ref{caszero}.

In Fig. \ref{caszero}, it is possible to see how the renormalized Casimir energy density $\mathcal{E}_0(a;r)$  and the Casimir pressure $\mathcal{P}_0(a;r)$ behave under the combined influence of the gravitational field and the curvature--coupling parameter $\xi$. Inside the neutron star ($r < R$), the solid curves exhibit a non-monotonic profile, where the energy density reaches a maximum near the stellar core and decreases toward the surface due to the gravitational redshift encoded in $e^{2\Phi(r)}$ and the radial dependence of the mass function $m(r)$. At the surface, the curves join smoothly to the Schwarzschild exterior (dashed region), where $m'(r) = 0$ and the purely gravitational modification dominates.
For both couplings, the Casimir energy gradually approaches the flat-space value (dotted lines) at large $r$.
The conformal case ($\xi = 1/6$) shows systematically higher magnitudes of $|\mathcal{E}_0|$, indicating that the curvature coupling amplifies the effective vacuum energy near the compact object.
This behavior reflects how spacetime curvature alters the zero-point modes contributing to the Casimir effect, producing a deviation from the standard Minkowski-space result that decays asymptotically as $r \to \infty$.

\begin{figure}[h!]
    \centering
    \includegraphics[width=0.45\linewidth]{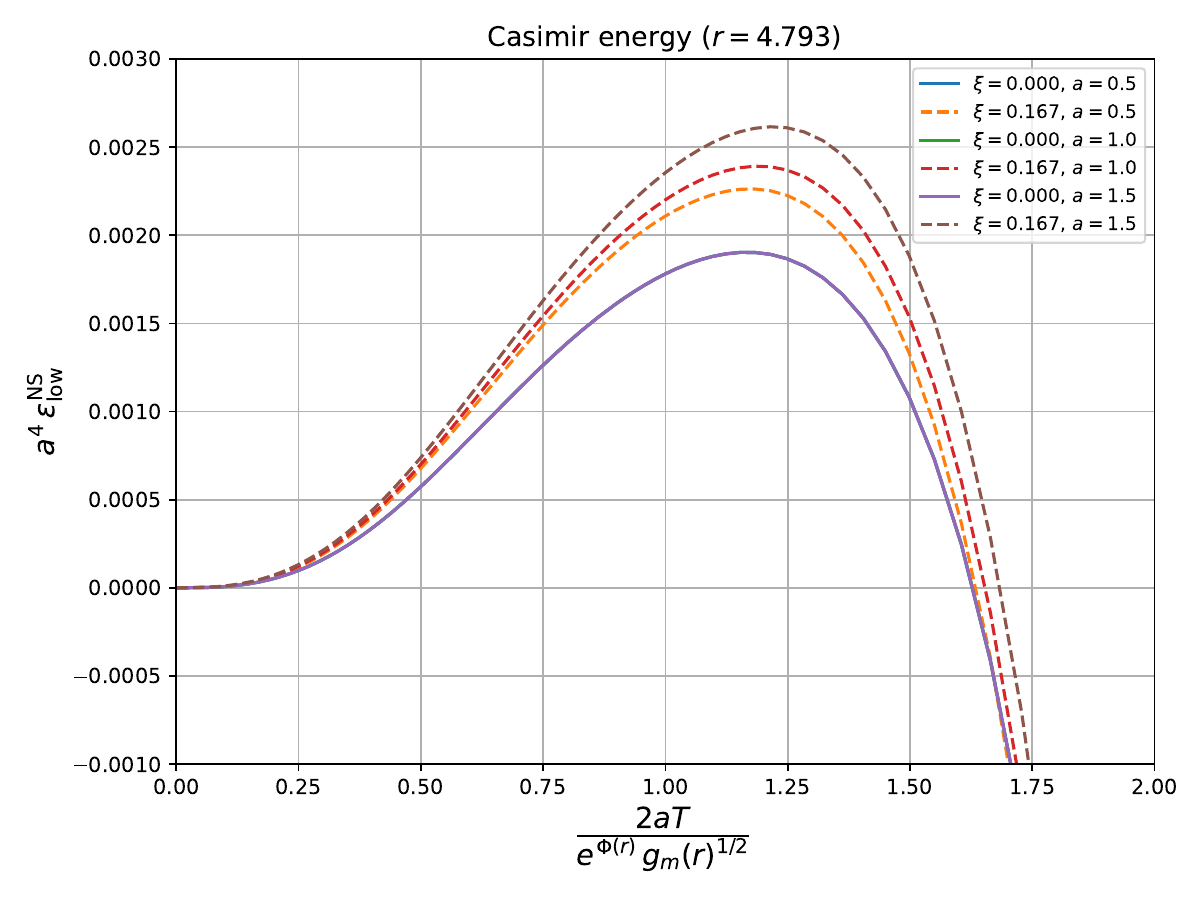}
    \includegraphics[width=0.45\linewidth]{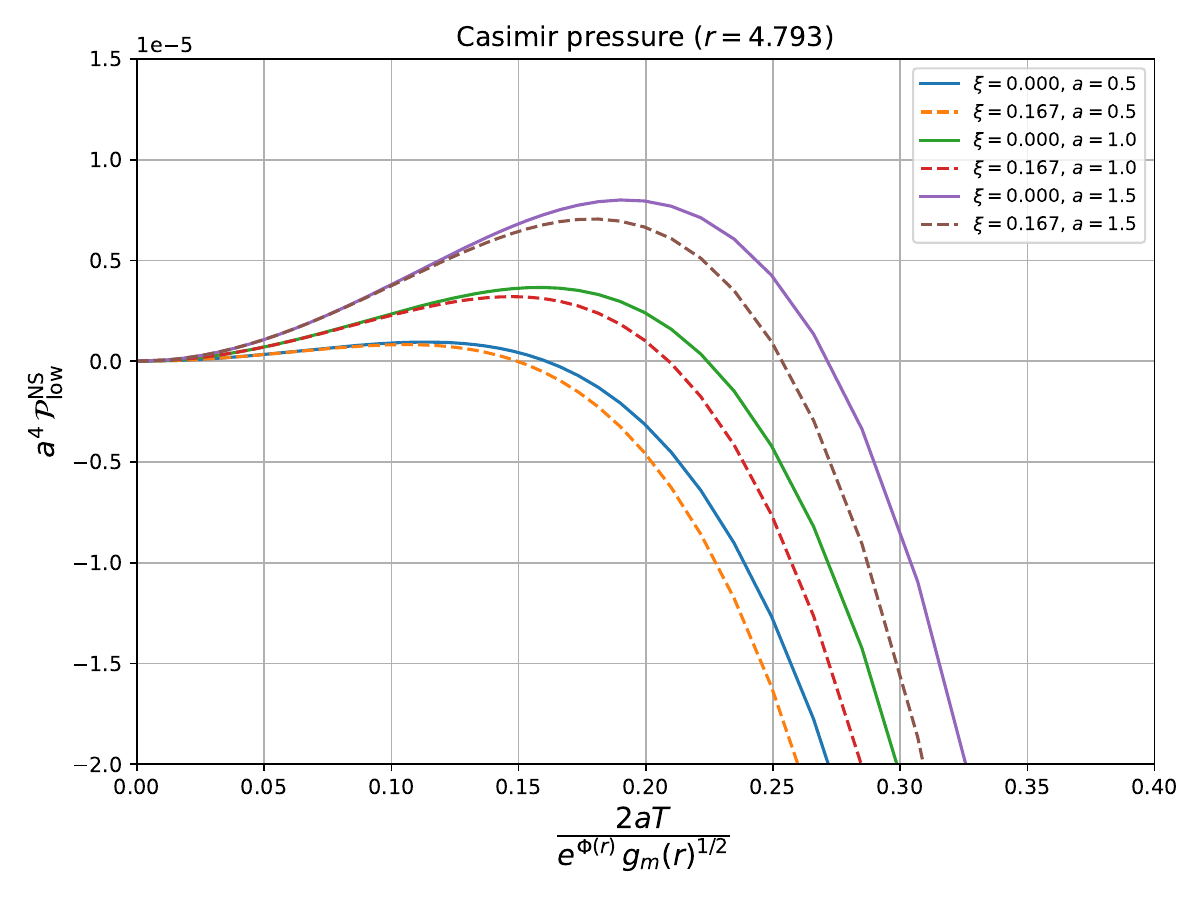}
    \caption{Thermal behaviour of the renormalized Casimir energy density (left panel) and radial pressure (right panel) inside the neutron star interior at a fixed radius $r = 4.793$. Both quantities are rescaled by $a^{4}$ and plotted as functions of the redshift–corrected combination $2aT\,e^{-\Phi(r)} g_{m}(r)^{-1/2}$, which is the natural local temperature parameter in the TOV background. }
    \label{figL}
\end{figure}

The behaviour of the dimensionless finite--temperature Casimir energy density \eqref{l1} and radial pressure \eqref{l4} inside the neutron--star interior is illustrated in Fig.~\ref{figL}, where both quantities are evaluated at the fixed radius $r = 4.793$.  To facilitate comparison between different plate separations, the observables are shown in dimensionless rescaled form
$a^{4}\varepsilon^{\rm NS}_{\rm low}$ and $a^{4}P^{\rm NS}_{\rm low}$, and are plotted as functions of the redshift--corrected combination $2aT\,e^{-\Phi(r)} g_{m}(r)^{-1/2}$, which represents the appropriate local temperature parameter in the TOV background.

The energy density increases from zero as the temperature rises, reaches a maximum, and subsequently decreases at high temperatures. This behaviour reflects the competition between thermal excitation of the modes and the geometric suppression introduced by the curved background. Larger values of the plate separation $a$ shift the peak to higher values of $2aT$, consistent with the fact that thicker cavities require higher thermal energy to excite the corresponding spectrum. The comparison between the two couplings $\xi = 0$ and $\xi = 1/6$ shows that the conformal value enhances the magnitude of the Casimir energy, demonstrating that the non--minimal coupling remains relevant even when the curvature at that radius is moderate.

The radial pressure exhibits a qualitatively similar behaviour: it grows positively at low temperature, reaches a maximum, and then becomes negative as the thermal contribution dominates. Overall, the figure shows that the combined effects of gravitational redshift, local curvature, thermal corrections, and the coupling $\xi$ significantly modify the Casimir observables within the neutron--star interior.

It is worth highlighting that in both cases, the renormalized energy density and the Casimir pressure diverge at high temperature $2aT\,e^{-\Phi(r)} g_{m}(r)^{-1/2}\gg1$, this behavior is expected since in a thermal Casimir system, it is usually not possible to describe the complete thermal spectrum with respect to the temperature on expression. For this reason, we should derive the thermal limits separately.

\begin{figure}[h!]
    \centering
    \includegraphics[width=0.45\linewidth]{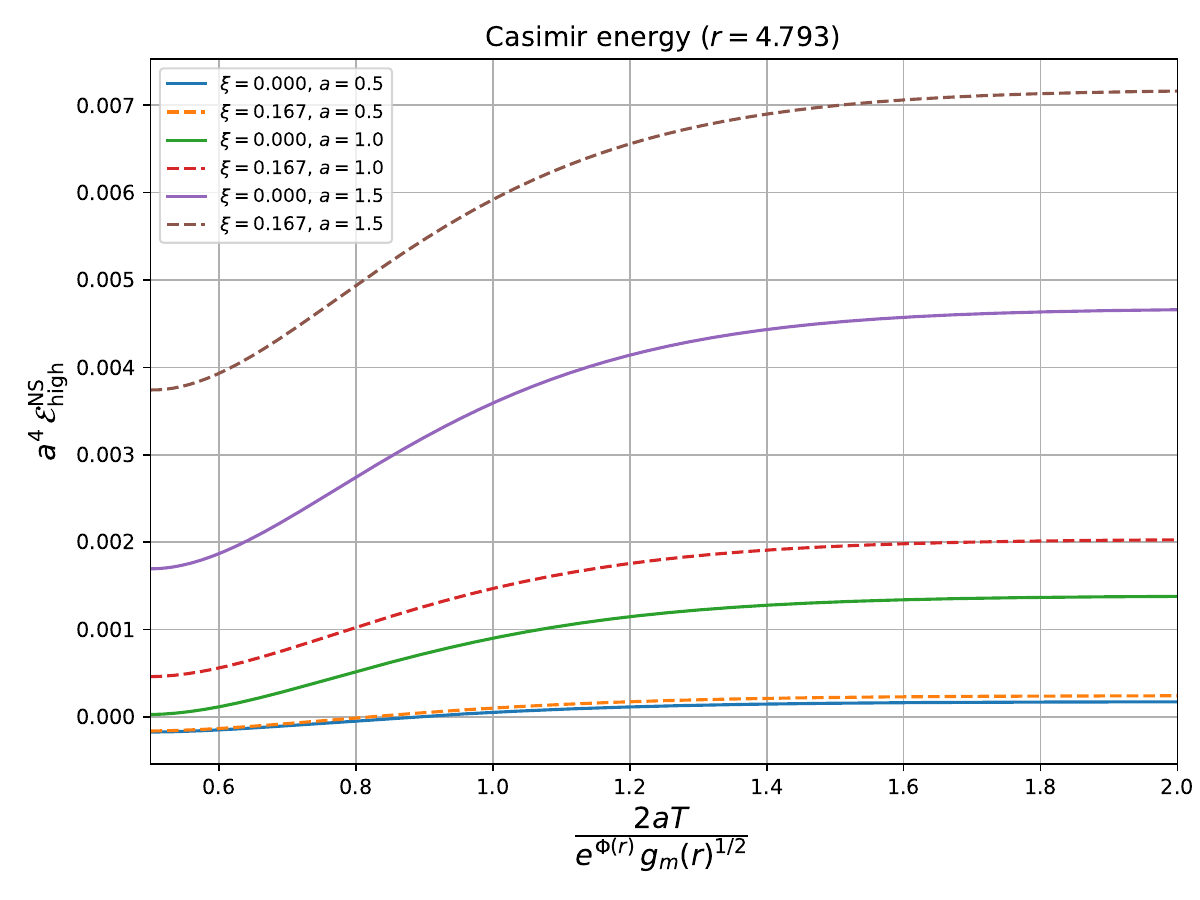}
    \includegraphics[width=0.45\linewidth]{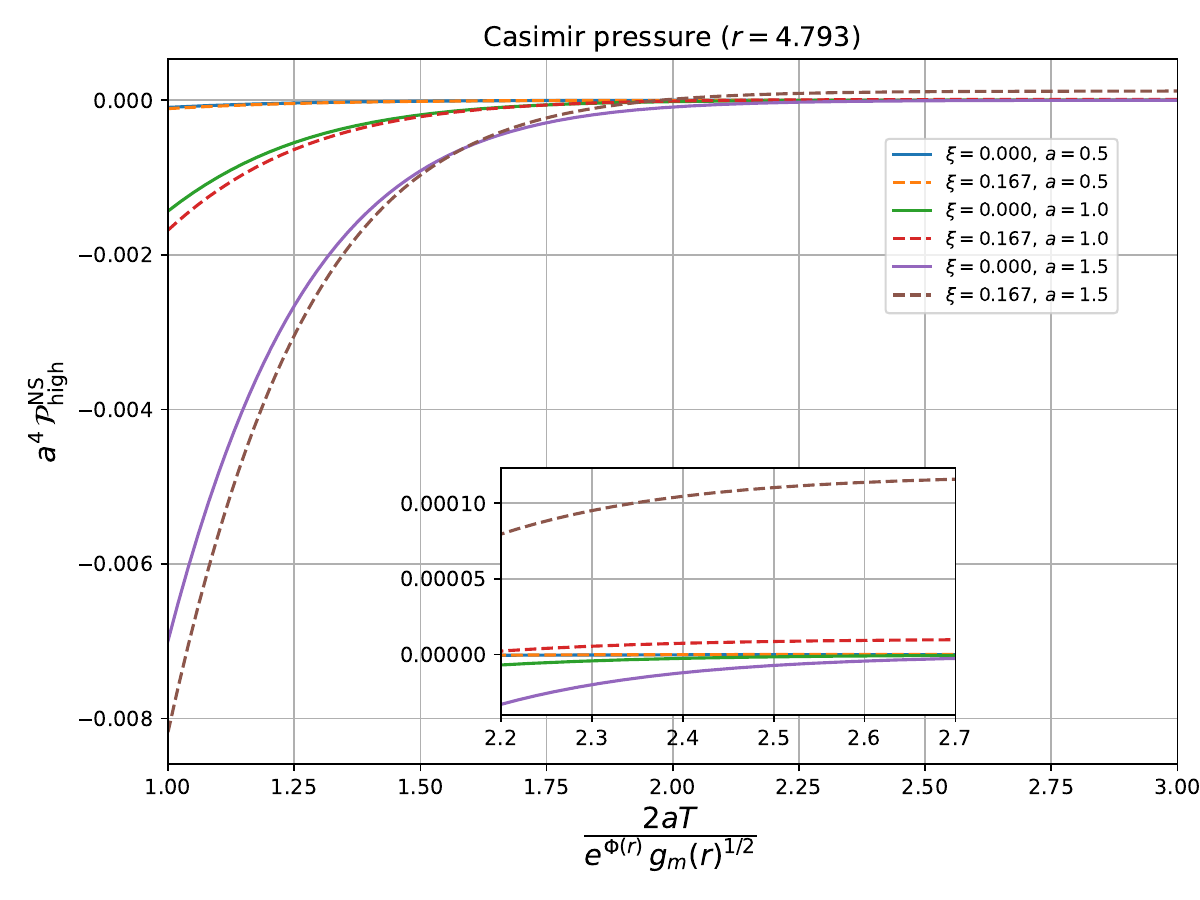}
    \caption{High-temperature behaviour of the renormalized Casimir energy density (left panel) and Casimir radial pressure (right panel) inside the neutron-star interior at the fixed radius $r = 4.793$. Both observables are rescaled by $a^{4}$ and plotted as functions of the redshift-corrected temperature parameter $2aT\,e^{-\Phi(r)} g_{m}(r)^{-1/2}$. }
    \label{figH}
\end{figure}

In the case of high temperature, for the Casimir energy and the Casimir radial pressure represented by Eqs. \eqref{h3} and \eqref{h6} respectively, the behaviour is shown in Fig. \ref{figH}.

Fig. \ref{figH} displays the high-temperature limit of the
renormalized Casimir energy density and radial pressure inside the neutron star, evaluated at the fixed radius $r = 4.793$.
Both plots use the rescaled quantities $a^{4}\varepsilon^{\rm NS}_{\rm high}$ and $a^{4}P^{\rm NS}_{\rm high}$ and are expressed in terms of the natural temperature variable $2aT\,e^{-\Phi(r)} g_{m}(r)^{-1/2}$, which incorporates the Tolman redshift and the effective radial metric factor.

It is worth highlighting that the analytic expression for
$P^{\rm NS}_{\rm high}$, Eq.~\eqref{h6}, contains terms that depend only on $\gamma_{r}$ and do not exhibit the exponential structure associated with spatial compactification (e.g., $e^{-2\pi/\gamma_{r}}$ or polylogarithmic terms). Such contributions correspond to the local Stefan–Boltzmann radiation rather than to the Casimir effect. For this reason, these non-Casimir thermal pieces must be removed in the numerical analysis so that the plotted quantity represents the properly renormalized Casimir pressure. After renormalization, the resulting curves clearly show that strong gravity, curvature, and the coupling $\xi$ significantly modify the asymptotic high-temperature behaviour of both the energy density and the pressure.

In contrast with the low-temperature regime, the high-temperature limit is monotonic: both the energy density and the pressure increase smoothly and approach constant finite values as $T \rightarrow \infty$, in agreement with the analytic form of the high-$T$ \eqref{h3} and \eqref{h7}. Larger plate separations yield larger asymptotic values, reflecting the dominant contribution of the thermal modes relative to the vacuum part. The comparison between $\xi = 0$ and $\xi = 1/6$ shows that the conformal coupling enhances the magnitude of both observables, with the effect becoming more pronounced for larger $a$.

The radial pressure becomes slightly negative at intermediate temperatures before tending to a small positive constant at very high $T$, a behaviour highlighted in the inset of the right panel. This sign change originates from the interplay between the geometric term and the thermal correction in the high-$T$ expansion. Overall, the figure demonstrates that gravity, the coupling $\xi$, and the plate separation all influence the asymptotic thermal regime, but the Casimir quantities remain finite and well behaved throughout, as expected from the renormalized high-temperature expressions.

The convergence of the system at high temperature, is expected since the quantum character of the system vanishes, which does not allow the Casimir energy or pressure to increase proportionally to the temperature. Moreover, at a low temperature in Fig. \ref{figH}, the behaviour presented is not adequate as it diverges or presents a non-physical, incompatible result.

\section{Conclusion}
\label{sec5}

In this work, we have investigated the thermal Casimir effect for a massless scalar field in the curved spacetime of a neutron star, employing the Thermo Field Dynamics (TFD) formalism. Starting from the renormalized energy–momentum tensor \eqref{eq2.13}, we extended the Stefan–Boltzmann law to include both the Tolman redshift factor and curvature corrections determined by the Tolman–Oppenheimer–Volkoff (TOV) metric, Eq. \eqref{eq2.24}, and compared it with limit results when $r>R$ is out of the Neutron star, Eq. \eqref{eq2.25}, and the flat spacetime $r\to\infty$, Eq. \eqref{eq2.26}. This approach provided a unified framework to describe vacuum effects in regions of intense gravitational fields, as shown by Eqs. \eqref{eq3.3} and \eqref{eq3.6}, which have the influence of the metric factor in the Casimir energy and pressure, respectively.

By introducing simultaneous temporal and spatial compactifications, we derived analytical expressions for the Casimir energy and pressure at finite temperature, covering the neutron star interior, the exterior Schwarzschild region, and the flat-space limits, Eqs. \eqref{1.38}--\eqref{1.43}. The resulting formulas consistently recover known results in the appropriate limits present in literature, confirming the robustness of the formalism.

Our analysis of the high- and low-temperature regimes, Eqs. \eqref{h3}-\eqref{h9} and \eqref{l1}-\eqref{l6}, revealed that curvature and redshift effects significantly modify the standard $T^4$ dependence of the energy density. In particular, strong gravity enhances the local energy density near the stellar surface while suppressing thermal contributions in the deep interior, indicating that quantum and gravitational effects are intimately coupled in such environments. The non-minimal coupling parameter $\xi$ introduces further modifications, acting as a measure of the deviation from minimal coupling and influencing the magnitude of the Stefan-Boltzmann law, Fig. \ref{nstar},  and also the sign of the Casimir energy and pressure, as demonstrated in the analysis considering the polytropic model applied to the TOV equations in Figs. \ref{caszero}-\ref{figH}.

These results emphasize that quantum vacuum fluctuations in curved spacetime are strongly affected by gravitational fields, especially within compact astrophysical configurations such as neutron stars. The framework presented here establishes a consistent basis for further extensions, including the study of massive or interacting fields, electromagnetic Casimir effects, and quantum corrections to neutron-star thermodynamics. Future work may also explore potential astrophysical implications, such as modifications to the thermal balance and equation of state due to vacuum energy contributions in strong-gravity regimes.


{\acknowledgments}

We would like to thank CNPq, CAPES and CNPq/PRONEX/FAPESQ-PB (Grant nos. 165/2018 and 015/2019), for partial financial support. K.E.L.F and R.A.B. would like to thank the Paraíba State Research Support Foundation FAPESQ  for financial support. M.A.A, F.A.B and E.P acknowledge support from CNPq (Grant nos. 306398/2021-4,
309092/2022-1, 304290/2020-3). A.R.Q work is supported by FAPESQ-PB. A.R.Q also acknowledges support by CNPq under process number 310533/2022-8.

\bibliography{bibli}

\appendix

\section{Neutron Star}
\label{Apx1}

To incorporate thermal effects into this spacetime, we employ the Thermo Field Dynamics (TFD) formalism. However, before proceeding, we first compute the Christoffel symbols associated with the metric \eqref{eq2.1}, defined by
\begin{equation}
\Gamma_{\mu\nu}^\rho = \frac{1}{2} g^{\rho\sigma} \left( \partial_\mu g_{\nu\sigma} + \partial_\nu g_{\mu\sigma} - \partial_\sigma g_{\mu\nu} \right).
\label{eq2.3}
\end{equation}

Due to the spherical symmetry of the system, many of the Christoffel symbols vanish, and the non-zero components are given below.

For the temporal components:
\begin{equation}
\Gamma^t_{tr} = \Gamma^t_{rt} = \Phi'(r),
\label{eq2.4}
\end{equation}
where $\Phi'(r) \equiv d\Phi/dr$. For the radial components:
\begin{subequations}\label{eq2.5.1}
\begin{align}
\Gamma^{r}_{tt} &= \Phi^{\prime}(r) e^{2 \Phi(r)}\left(1 - \frac{2Gm(r)}{r}\right), \label{eq2.5.1} \\[4pt]
\Gamma^{r}_{rr} &= -\frac{G\left[m(r) - r m^{\prime}(r)\right]}{r^2 - 2G r m(r)}, \label{eq2.5.2} \\[4pt]
\Gamma^{r}_{\theta\theta} &= 2Gm(r) - r, \label{eq2.5.3} \\[4pt]
\Gamma^{r}_{\phi\phi} &= \sin^2\theta \left[2Gm(r) - r\right]. \label{eq2.5.4}
\end{align}
\end{subequations}

and for the angular components:
\begin{subequations}\label{eq:Gamma_ang_components}
\begin{align}
\Gamma^{\theta}_{r\theta} &= \Gamma^{\theta}_{\theta r} = \frac{1}{r}, \label{eq2.6.1} \\[4pt]
\Gamma^{\phi}_{r\phi} &= \Gamma^{\phi}_{\phi r} = \frac{1}{r}, \label{eq2.6.2} \\[4pt]
\Gamma^{\theta}_{\phi\phi} &= -\sin\theta \cos\theta, \label{eq2.6.3} \\[4pt]
\Gamma^{\phi}_{\theta\phi} &= \Gamma^{\phi}_{\phi\theta} = \cot\theta. \label{eq2.6.4}
\end{align}
\end{subequations}

The Ricci tensor is obtained from the Riemann tensor $R_{\sigma \mu \nu}^\rho$ via contraction $R_{\mu \nu}=R_{\mu \rho \nu}^\rho$, and the non-zero components of the Ricci tensor are

\begin{subequations}\label{eq2.7.1}
\begin{align}
R_{tt} &= e^{2\Phi(r)}\left\{[\Phi''(r)+\Phi'(r)^2]\left(1-\frac{2Gm(r)}{r}\right)
+ \Phi'(r)\left[\frac{2r - Grm'(r) - 3Gm(r)}{r^2}\right]\right\}, \label{eq2.7.1} \\[4pt]
R_{rr} &= -\Phi''(r) - \Phi'(r)^2 + \frac{G[r m'(r) - m(r)] [r\Phi'(r) + 2]}{r^2 [r - 2Gm(r)]}, \label{eq2.7.2} \\[4pt]
R_{\theta\theta} &= [2Gm(r) - r]\Phi'(r) + G\!\left[m'(r) + \frac{m(r)}{r}\right], \label{eq2.7.3} \\[4pt]
R_{\phi\phi} &= \sin^2\theta\,R_{\theta\theta}. \label{eq2.7.4}
\end{align}
\end{subequations}

Now, we should compute the Ricci Scalar $R$, which is obtained from the contraction of the Ricci tensor with the inverse metric:
\begin{equation}
    R=g^{\mu \nu} R_{\mu \nu}=-e^{-2 \Phi(r)} R_{t t}+\left(1-\frac{2 G m(r)}{r}\right) R_{r r}+\frac{1}{r^2} R_{\theta \theta}+\frac{1}{r^2 \sin ^2 \theta} R_{\phi \phi}.
    \label{eq2.8}
\end{equation}

Substituting the Ricci tensor components, \eqref{eq2.7.1}-\eqref{eq2.7.4}, in eq. \eqref{eq2.8}, we get:
\begin{equation}
    R=2\left\{\frac{\Phi^{\prime}(r)}{r^2}\left[3 G m(r)+r G m^{\prime}(r)-2 r\right]-\left[\Phi^{\prime \prime}(r)+\Phi^{\prime}(r)^2\right]\left[1-\frac{2 G m(r)}{r}\right]\right\}
    \label{eq2.9}
\end{equation}

\section{TOV equations for a static, spherically symmetric star}
\label{Apx2}

We consider a static, spherically symmetric spacetime with the line element
\begin{equation}
ds^2 = -e^{2\Phi(r)} dt^2 + \left(1 - \frac{2Gm(r)}{r}\right)^{-1} dr^2 + r^2 d\Omega^2,
\label{eqB2.1}
\end{equation}
Considering a model where the matter is a perfect fluid with four-velocity $u^\mu = e^{-\Phi}\,\delta^\mu_{\ t}$, the stress-energy tensor is expressed as
\begin{equation}
T^{\mu\nu} = (\epsilon + P)\,u^\mu u^\nu + P\,g^{\mu\nu},
\label{eqB2.2}
\end{equation}
with $\epsilon(r)$ the total energy density and $P(r)$ the pressure.

From the Einstein equations, the mass accumulation relation contained in a radius $r$ is given by
\begin{equation}
\frac{dm}{dr} = 4\pi r^2\,\epsilon(r).
\label{eqB2.3}
\end{equation}
From the energy-momentum conservation $\nabla_\mu T^{\mu\nu}=0$, the following relation can be obtained
\begin{equation}
\frac{dP}{dr} = -(\epsilon+P)\,\frac{d\Phi}{dr}.
\label{eqB2.4}
\end{equation}
where
\begin{equation}
    \frac{d P}{d r}=-\frac{(\epsilon+P)\left(m+4 \pi r^3 P\right)}{r(r-2 m)}\,,
    \label{eqB2.5}
\end{equation}
The set of equations given by \eqref{eqB2.3}-\eqref{eqB2.5} is the TOV equation.

Writing $\nu(r)\equiv 2\Phi(r)$ for convenience, it can be determined from
\begin{equation}
    \frac{d \nu}{d r}=-\frac{2}{\epsilon+P} \frac{d P}{d r},
    \label{eqB2.6}
\end{equation}
which leads us to
\begin{equation}
\frac{d\nu}{dr} = \frac{2G\!\left( m + 4\pi r^3 P \right)}{r^2\!\left(1 - \dfrac{2Gm}{r}\right)}.
\label{eqB2.7}
\end{equation}

The numerical evaluation is obtained by considering some values from the center, i.e., $r=0$, which leads us to
\begin{equation}
P(0)=P_c,\qquad m(0)=0,\qquad \nu(0)=-1,
\label{eqB2.7}
\end{equation}
where $\nu$ is arbitrary and will be fixed by matching to the exterior solution at the stellar surface $r=R$

In the case of the near center, we should do expansions.
Assuming regularity and denoting $\epsilon_c\equiv \epsilon(0)$, the series expansions read
\begin{align}
P(r) &\simeq P_c - 2\pi\,(\epsilon_c + P_c)\!\left(P_c + \frac{\epsilon_c}{3}\right) r^{2} + \mathcal{O}(r^{4}),
\label{eqB2.8a}\\
m(r) &\simeq \frac{4\pi}{3}\,\epsilon_c\, r^{3} + \mathcal{O}(r^{5}),
\label{eqB2.8b}\\
\Phi(r) &\simeq \Phi_c + 2\pi\!\left(P_c + \frac{\epsilon_c}{3}\right) r^{2} + \mathcal{O}(r^{4}).
\label{eqB2.8c}
\end{align}

On the surface of the neutron star where $r\ge R$ the spacetime is Schwarzschild with total gravitational mass $M\equiv m(R)$. Continuity of $g_{tt}$ at $r=R$ fixes
\begin{equation}
\nu_\star=-\lambda_\star=\ln\!\left(1-\tfrac{2M}{R}\right)\,,
\label{eqB2.9}
\end{equation}
with $e^{\lambda}=(1-2Gm/r)^{-1}$.

The gravitational mass may be written as
\begin{equation}
M = m(R) = \int_{0}^{R} 4\pi r^{2}\,\epsilon(r)\,dr,
\label{eqB2.10}
\end{equation}
or, using the Tolman mass formula,
\begin{equation}
M = \int_{0}^{R} 4\pi r^{2}\,e^{(\nu+\lambda)/2}\,\big(\epsilon + 3P\big)dr\,.
\label{eqB2.11}
\end{equation}
The baryonic mass stan is given by
\begin{equation}
M_0 = \int_{0}^{R} 4\pi r^{2}\,e^{\lambda/2}\,\rho(r)\,dr,
\label{eqB2.12}
\end{equation}

\end{document}